\documentclass[usenatbib]{mnras}

\usepackage[T1]{fontenc}
\usepackage{ae,aecompl}

\usepackage{amssymb}	% Extra maths symbols
\usepackage{natbib}
\usepackage{graphicx}
\usepackage{rotating,times,graphicx,latexsym}
\usepackage{color}
\usepackage{longtable}
\usepackage{amsmath}
\usepackage{lscape}
\usepackage{array}
\usepackage{soul} % To add highlighting of text
\usepackage[pdftex,dvipsnames]{xcolor}  % Coloured text etc.
\usepackage{amsmath}
\newcommand{\dg}{$^{\circ}$}

\newcommand{\hc}{{\it HOYS-CAPS}}
\newcommand*\samethanks[1][\value{footnote}]{\footnotemark[#1]}
\title[HOYS-CAPS]{A survey for variable young stars with small telescopes: First results from HOYS-CAPS}

\author[Froebrich et al.]{\parbox{\textwidth}{
D.~Froebrich$^{1}$\thanks{E-mail: df@star.kent.ac.uk},
J.~Campbell-White$^{1}$,
A.~Scholz$^{2}$,
J.~Eisl\"{o}ffel$^{3}$,
T.~Zegmott$^{1}$\thanks{Observer Beacon Observatory},
S.J.~Billington$^{1}$\samethanks,
J.~Donohoe$^{1}$\samethanks,
S.V.~Makin$^{1}$\samethanks,
R.~Hibbert$^{1}$\samethanks,
R.J.~Newport$^{4}$\samethanks,
R.~Pickard$^{5}$\thanks{HOYS-CAPS Observer},
N.~Quinn$^{5}$\samethanks,
T.~Rodda$^{5}$\samethanks,
G.~Piehler$^{6}$\samethanks,
M.~Shelley$^{7}$\samethanks,
S.~Parkinson$^{5}$\samethanks,
K.~Wiersema$^{8,9}$\samethanks,
I.~Walton$^{5,10}$\samethanks
 \vspace{0.4cm}} \\
$^1$ Centre for Astrophysics and Planetary Science, School of Physical Sciences, University of Kent, Canterbury, CT2 7NH, UK \\
$^2$ SUPA, School of Physics and Astronomy, University of St Andrews, North Haugh, St Andrews KY16 9SS, UK \\
$^3$ Th\"uringer Landessternwarte, Sternwarte 5, 07778 Tautenburg, Germany\\
$^4$ Functional Materials Group, School of Physical Sciences, University of Kent, Canterbury, CT2 7NH, UK \\
$^5$ The British Astronomical Association, Variable Star Section, Burlington House Piccadilly, London W1J 0DU, UK \\
$^6$ Selztal Observatory, D-55278 Friesenheim, Bechtolsheimer Weg 26, Germany \\
$^7$ Ashford Astronomical Society, Woodchurch Memorial Hall, Woodchurch, TN26 3QB, UK \\
$^8$ Department of Physics and Astronomy, University of Leicester, University Road, Leicester, LE1 7RH, UK \\
$^9$ Department of Physics, University of Warwick, Coventry, CV4 7AL, UK \\
$^{10}$ Cranbrook and District Science and Astronomy Society, Cranbrook School, Waterloo Road, Cranbrook, TN17 3JD, UK \\
}

\begin{document}

\date{Received sooner; accepted later}
\pagerange{\pageref{firstpage}--\pageref{lastpage}} \pubyear{2017}
\maketitle

\label{firstpage}

\begin{abstract}

Variability in Young Stellar Objects (YSOs) is one of their primary characteristics. Long-term, multi-filter, high-cadence monitoring of large YSO samples is the key to understand the partly unusual light-curves that many of these objects show. Here we introduce and present the first results of the \hc\ citizen science project which aims to perform such monitoring for nearby ($d$\,$<$\,1\,kpc) and young (age\,$<$\,10\,Myr) clusters and star forming regions, visible from the northern hemisphere, with small telescopes. We have identified and characterised 466 variable (413 confirmed young) stars in 8 young, nearby clusters. All sources vary by at least 0.2\,mag in V, have been observed at least 15 times in V, R and I in the same night over a period of about 2\,yrs and have a Stetson index of larger than 1. This is one of the largest samples of variable YSOs observed over such a time-span and cadence in multiple filters. About two thirds of our sample are classical T-Tauri stars, while the rest are objects with depleted or transition disks. Objects characterised as bursters show by far the highest variability. Dippers and objects whose variability is dominated by occultations from normal interstellar dust or dust with larger grains (or opaque material) have smaller amplitudes. We have established a hierarchical clustering algorithm based on the light-curve properties which allows the identification of the YSOs with the most unusual behaviour, and to group sources with similar properties. We discuss in detail the light-curves of the unusual objects V2492\,Cyg, V350\,Cep and 2MASS\,J21383981$+$5708470.

\end{abstract}


\section{Introduction}

Time-domain observations of star forming regions are a reliable source of information about the formation and early evolution of stars. Historically, young stars were first discovered based on their irregular and large-amplitude optical variability \citep{1945ApJ...102..168J}. Starting in the late 1980s, the prevalent rotational flux modulation observed in young stars has been used to measure thousands of rotation periods ranging from hours to weeks, a great foundation for studies of angular momentum evolution during the protostellar stages and beyond (see Protostars and Planets reviews by \citet{2007prpl.conf..297H}, \citet{2014prpl.conf..433B}). In addition to rotation, optical fluxes of young stars are affected by variable excess emission from accretion shocks, variable emission from the inner disk, and variable extinction along the line of sight \citep{2001AJ....121.3160C}, and can therefore give insights into the structure and evolution of the environment of young stellar objects (YSOs).

While the interplay of these variability causes can lead to very complicated light-curves and render the interpretation difficult, several prototypical phenomena have been successfully attributed to a physical cause. AA\,Tau is now the prototype for a category of 'dippers', a contingent of young stars temporarily eclipsed by portions of the inner disks which are warped by the star's magnetic field \citep{2014prpl.conf..433B,2015A&A...577A..11M}. FU\,Ori and EX\,Lupi are prototypes for stars which experience sharp increases in their mass accretion rates \citep{2014prpl.conf..387A}, a phenomenon that is now known to occur on a wide range of timescales \citep{2016AJ....151...60S}, including objects with continuous accretion rate changes and consequently stochastic light-curves \citep{2014AJ....147...83S,2016AJ....151...60S}. While the detailed physics behind the variety of accretion bursts is still under debate, the episodic and unstable nature of accretion is now established as a primary characteristic of the early stellar evolution. Many variable young stars still defy classification and are complicated to understand (e.g., the recent dimmings of RW\,Aur, see \citet{2016MNRAS.463.4459B}).

The 'gold standard' for optical studies of YSO variability are space-based observing campaigns with COROT and Kepler/K2. The combined COROT/Spitzer monitoring of NGC2264 is unprecedented in cadence, multi-wavelength coverage, and photometric precision. Complemented by ground-based observations, it has led to a new comprehensive overview about the phenomenology of young variable stars and the underlying causes \citep{2014AJ....147...82C}. Kepler/K2 has observed large numbers of young stars continuously over campaigns of 70\,d; its archive is a treasure trove for detailed studies of rotation periods, dippers, bursters, and related phenomena \citep{2016ApJ...816...69A}. 

So far, these studies have been mostly limited to individual regions, to baselines of weeks to months, or to one optical band. While the astrometry mission Gaia will add time-domain information over 5\,yr for thousands of young stars, it will only provide sparse cadence. The same is true for the LSST coverage. There is a definite need for long-term monitoring, quasi-simultaneous in multiple bands, similar to the pioneering studies by \citet{2007A&A...461..183G}, but extending to a large and unbiased sample of young stars, to capture the full range of YSO variability and to put the known phenomena in context. This can be done from the ground and with modestly sized telescopes. 

This is the goal of the \hc\footnote{\tt \href{http://astro.kent.ac.uk/~df/hoyscaps/index.html}{http://astro.kent.ac.uk/$\sim$df/hoyscaps/index.html}} project, presented for the first time in this paper. Primarily we publish a large catalog of variable young stars in regions across the northern hemisphere, including many clusters previously unstudied on long timescales. For all variable stars, we derive a range of metrics that are used to classify the stars and to provide clues about the origin of the variability. So far, the survey covers baselines up to 2 years. This project will continue to extend the time window and will provide a foundation for detailed studies of specific phenomena.

This paper is organised as follows. In Sect.\, \ref{data} we introduce the \hc\ project and present details of the utilised observatories and data calibration. Section\,\ref{dataandanalysis} presents the selection of young clusters and YSOs and discusses the data analysis procedures applied to their light-curves. Finally, in Sect.\,\ref{results} we discuss the results of our analysis.

\section{Data}\label{data}

\subsection{The \hc\ Project}\label{hc_project}

\hc\ stands for {\it Hunting Outbursting Young Stars with the Centre of Astrophysics and Planetary Science}. This citizen science project has been run by the University of Kent since October 2014. It currently involves amateur astronomers from the UK, as well as from Europe. It is also supported by additional professional observatories (see Sect.\,\ref{other_observatories} for observatory details). The participants take images of objects on our target list, perform a basic data reduction (dark/bias and flat-field correction) and submit these reduced images for inclusion into our database via our web-interface\footnote{\tt \url{http://astro.kent.ac.uk/HOYS-CAPS/}}. 

The aim of \hc\ is the long term, multi-filter optical photometric monitoring of young (age less than 10\,Myr), nearby (distances typically within 1\,kpc) star clusters or star forming regions visible from the northern hemisphere with small telescopes. There are no restrictions given to the participants in terms of observing cadence, target priority, field of view, integration times or filter selection. 

At the time of writing, the \hc\ target list contains 17 young clusters/regions as well as several additional targets selected from the Gaia Photometric Alerts\footnote{\tt \url{http://gsaweb.ast.cam.ac.uk/alerts/alertsindex}}, some of which are within the \hc\ target regions. In total almost 3200 images have been taken for the project, with a total of almost 970\,hrs of observing time. About 80\,\% of the \hc\ images (corresponding to 90\,\% of the observing time) have been obtained with the Beacon Observatory.

\subsection{The Beacon Observatory}\label{sect_beacon}

The majority of the data presented in this paper has been taken by post-graduate student observers at the University of Kent's Beacon Observatory.
The Beacon Observatory consists of a 17\arcsec\ {\em Planewave} Corrected Dall-Kirkham (CDK) Astrograph telescope situated at the University of Kent (51.296633\dg\ North, 1.053267\dg\ East, 69\,m elevation). The telescope is equipped with a 4k\,$\times$\,4k Peltier-cooled CCD camera and a B, V, R, I, H$\alpha$ filter set. The pixel scale of the detector is 0.956\arcsec, giving the camera a field of view of about 1\dg\,$\times$\,1\dg. Due to the optical system of the telescope the corners of the detector are heavily vignetted. Hence the usable field of view of the detector is a circular area with a diameter of approximately 1\dg.

The observatory has, despite its location, a good record for observations. Over the first two years of operations an average of 10 nights per month were used for science observations, with an average of 50\,hrs per month usable, i.e. just above 50\,\% of the time is used in each night with clear skies. The typical seeing in the images is about 3\arcsec\,--\,4\arcsec.

Images taken by the observatory for the \hc\ project are typically taken in the following sequence: 120\,s integrations are done in V, R, and I and this sequence is repeated 8 times. Including filter changes and CCD readout, this sequence takes one hour. All individual images are dark and bias subtracted and flat-fielded using sky-flats. All images taken of a particular target during a sequence are median averaged using the Montage software package\footnote{\tt \url{http://montage.ipac.caltech.edu/}}. 

\subsection{Details on additional Observatories used}\label{other_observatories}

Here we give a brief description of the other observatories and telescopes used to obtain data for the \hc\ project.

{\bf Selztal-Observatory:} The observatory is located in Friesenheim, approximately 20\,km South of Mainz in Germany. The telescope is a 20\arcsec\ Newton, with f\,=\,2030\,mm and an ASA corrector and an ASA DDM 85 Pro mount. The CCD used is a STL 11000M with anti-blooming gate and a set of RGB filters is available. Twilight flats are taken to correct for variations in pixel sensitivity and image processing is performed with the Maxim DL software. Typical exposure times are 120\,--\,300\,s and observations are guided with an accuracy of less than one pixel and seeing of about 3\arcsec. Due to surrounding street lights, there are some gradients left in the images not corrected for by the flat- field, but they do not influence the photometry.

{\bf Ponteland Observatory:} The observatory is based in Ponteland (about 10\,km North-West of Newcastle upon Tyne, UK) at 55.0525\dg\ North, 1.73889\dg\ West. The telescope used is a  235 mm SCT (f/6.3) with an Atik460 mono camera and Bessel B, V, R, I, C filters. This provided a plate scale of approx 0.63\arcsec/pixel and a field of view of 29\arcmin\,$\times$\,23\arcmin. Typical seeing in the images is around 3\arcsec\ and exposures times range from 30\,s to 120\,s depending on target and filter.

{\bf Steyning Observatory:} The observatory is situated in Steyning, West Sussex, UK. The telescope is an 8\arcsec (200\,mm) Ritchey Chretien (f/8.0) operating at a focal length of 1600\,mm with a Santa Barbara Instrument Group (SBIG) STF-8300M mono camera, and a 'green' filter from a tri-colour imaging set made by Astronomik. Using 2\,x\,2 binned pixels, this provides a plate scale of about 1.4\arcsec/pixel with a field of view of 39\arcmin\,$\times$\,29\arcmin. Integration times for the images range from 60\,s to 240\,s. Image calibration (darks, flat-fields and stacking) is carried out with the AstroArt software.

{\bf Piers Sellers Observatory:} The observatory is run by the Cranbrook and District Science and Astronomy Society (CADSAS) situated in Cranbrook in Kent -- about 20\,km East of Royal Tunbridge Wells. The telescope is the 0.57\,m 'Alan Young' f/4.7 Newtonian reflector which uses a ZWO ASI 174\,MM (mono-cooled) camera at prime focus. Typically images are taken as 10\,$\times$\,10\,s integrations and are co-added. Calibration is performed using the Deep Sky Stacker\footnote{\tt \url{http://deepskystacker.free.fr/english/index.html}} or AIP4WIN software \citep{2005haip.book.....B}.

{\bf Astcote Observatory:} The observatory is situated about 15\,km South-West of Northampton. The telescope is a C9.25\arcsec\ (f/6.3) with a Starlight Xpress MX916 CCD and EQ6 mount. Typically images are taken with a V-Band filter with 35\,s integration time. Basic calibrations with darks and sky flats are performed using the AIP4Win image processing software.

{\bf High Halden Observatory:} This observatory situated 15\,km South-West of Ashford, Kent. The telescope is a Takahashi Epsilon 180ED  8\arcsec\ (f/2.8) hyperbolic Newtonian. The camera used is a Canon 350D digital single-lens reflex camera (DSLR) with all internal filters removed and a Baader IR/UV cut filter is used in conjunction. Integration times for the images range from 80\,min to 180\,min in 5\,min sub-exposures. Image calibration (flats, darks, bias and stacking) is performed with Christian Buil's IRIS software.

{\bf Shobdon Observatory:} The observatory is situated in Herefordshire about 8\,km from the UK/Wales Border. It houses a Meade LX200 35\,cm SCT (f/7.7) operating at a focal length of 2500\,mm with a Starlight XPress SXV-H9 CCD and a set of Johnson-Cousins B, V, R and I filters. Integration times are typically 60\,s and darks and flats are applied using AIP4WIN software.   

%{\bf University of Leicester Observatory:} The University of Leicester runs a 0.5\,m telescope (UL50). This is a 20\arcsec\ Planewave CDK telescope with a SBIG ST2000XM CCD camera. It is equipped with a Johnson-Cousins B, V, R, I filter-set. Data were reduced using dark, bias and flat-frames taken the same night, using an {\sc IRAF} pipeline.

{\bf University of Leicester Observatory:} The University of Leicester runs a 0.5\,m telescope (the University of Leicester 50cm, or UL50). This is a 20\arcsec\ {\em Planewave} CDK telescope with a SBIG  ST2000XM camera.  It is equipped with a Johnson-Cousins B, V, R, I filter-set. Data were reduced using dark, bias and flat-frames taken the same night as science observations, using an {\sc IRAF} pipeline.

{\bf Thuringian State Observatory:} The Th\"uringer Landessternwarte is operating its Alfred-Jensch 2-m telescope\footnote{\tt \url{http://www.tls-tautenburg.de/TLS/index.php?id=25&L=1}} near Tautenburg (50.980111\dg\ North, 11.711167\dg\ East, 341\,m elevation). For HOYS-CAPS the telescope is used in its Schmidt configuration (clear aperture 1.34\,m, mirror diameter 2.00\,m, focal length 4.00\,m). It is equipped with a 2k\,$\times$\,2k liquid nitrogen-cooled CCD camera and with a B, V, R, I, H$\alpha$ filter set. The employed SITe CCD has 24\,$\mu$m\,$\times$\,24\,$\mu$m pixels, leading to a field of view of 42\arcmin\,$\times$\,42\arcmin.  Single exposures of 20 to 120\,s integration time -- depending on the filter -- are obtained, and several consecutive frames may be co-added. Dark frames and dome-flats are used for image calibration.

{\bf LCO telescopes:} In addition, some of the amateur astronomers used access to the range of telescopes from the Las Cumbres Observatory (LCO). LCO provides a range of 2\,m, 1\,m and 0.4\,m telescopes located at various sites around the Earth to allow complete longitudinal coverage. The two 2\,m telescopes are the Faulkes telescopes built by Telescope Technologies Ltd. which are f/10 Ritchey-Cretien optical systems. The 1\,m telescopes are also Ritchey-Cretien systems with f/7.95, while the 0.4\,m telescopes are Meade 16\arcsec\ RCX telescopes. Data included in this work has been taken on Haleakala Observatory (0.4\,m, 2\,m), Siding Spring Observatory (0.4\,m, 1\,m) and Tenerife (0.4\,m). All data from LCO are returned reduced with dark and flat-field corrections applied. Integration times are typically 60\,s but depend on the target and telescope size.

\subsection{General data calibration}

The astrometric solutions for all images obtained for the \hc\ project are determined using the {\tt astrometry.net} software \citep{2008ASPC..394...27H}. Photometry for all images is performed using the Source Extractor \citep{1996A&AS..117..393B}. 

As indicated above, the data included in the analysis for this paper comes mostly from the Beacon Observatory but is supported by a variety of other telescopes. Hence, a mix of CCD cameras and DSLRs was used to capture these supporting images. All data included in the analysis of this paper has either been taken in a standard Johnson or Cousins optical or I-band filter or as part of a tri-colour RGB filter from a DSLR. 

\begin{figure}
\centering
\includegraphics[angle=0,width=\columnwidth]{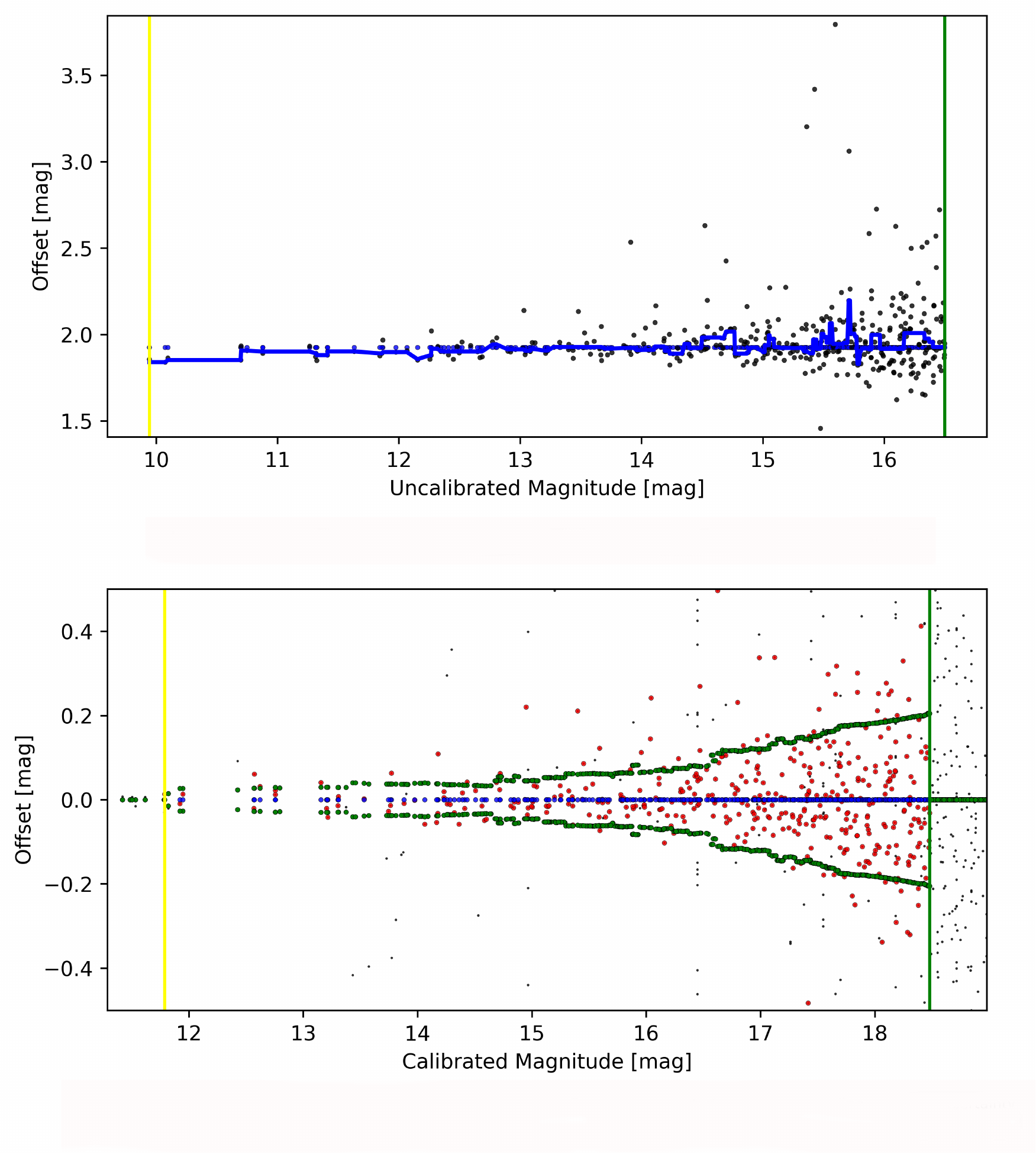} 
\caption{Example of the calibration of an image of IC\,348 in a green (TG) filter into the V filter. {\bf Top panel:} Difference {\bf ($m^i-m^r$)} of the instrumental magnitudes between the TG image and the V-Band reference dataset. The blue line indicates the running median of the data-points and the two vertical lines indicate the range of instrumental magnitudes considered. {\bf Bottom panel:} Difference {\bf ($m-m^r$)} between the calibrated TG data and the V-Band reference frame. The larger (red) dots indicate stars with no problems in the photometry, while the remaining stars are shown as smaller (black) dots. The dotted (green) data points indicate the one sigma scatter for stars within 0.5\,mag and the two vertical lines the range of magnitudes in which stars are included in the calibration. \label{fig_example_calibration}}
\end{figure}

All photometry obtained from each image has been calibrated relative to one of the Beacon Observatory datasets. Hence, for each target (see Sect.\,\ref{yso_selection}) and filter (VRI) we identified a deep image taken under photometric conditions as instrumental magnitude reference frame. All other magnitudes from all other images have been calibrated relative to these references. This has been done by matching all stars in the images to the respective reference frame. Only stars detected with a flag of '0' from the Source Extractor (see \citet{1996A&AS..117..393B} for details), indicating no problems with the photometry, have been used to obtain the calibration equation. To convert the instrumental magnitudes ($m^i$) of each image into the calibrated instrumental magnitudes ($m$) of the reference frame the following steps are performed: i) For all stars determine the magnitude difference ($\Delta m = m^i - m^r$) between the instrumental magnitudes and the magnitudes in the reference frame ($m^r$) and plot against $m^i$ (see top panel in Fig.\,\ref{fig_example_calibration}); ii) Using a least-square optimization find the best fitting eight parameters for the function $f(m^i)$ that minimize $m^r - f(m^i)$. Equation\,\ref{eq_phot} shows the parameterisation of $f(m^i)$, where $\mathcal{P}_4(m^i)$ denotes a 4$^{\rm th}$ order polynomial and the other term is a photocurve function proposed by \citet{2005MNRAS.362..542B} and \citet{1969A&A.....3..455M}; iii) Determine the calibrated magnitudes for all stars using $m = f(m^i)$; iv) Plot the difference $m - m^r$ against the calibrated magnitudes to check if the calibration has been successful (see bottom panel in Fig.\,\ref{fig_example_calibration}).

\begin{equation}\label{eq_phot}
f(m^i) = A \cdot \log \left( 10^{B \cdot (m^i-C)} + 1 \right) + \mathcal{P}_4(m^i)
\end{equation}

Typically the calibration has an accuracy of a few percent for stars brighter than 14$^{\rm th}$ or 15$^{\rm th}$ magnitude, which rises to 0.2\,mag for the faintest detected stars. See Fig.\,\ref{fig_example_calibration} for an example of how the calibration works for data from IC\,348 taken in a green (TG) filter at the Seltztal-Observatory when calibrated into the V-Band reference data. When the calibration has not resulted in these typical uncertainties (less than 0.2\,mag scatter for the faintest objects and less than 0.05\,mag for the brighter stars), the images have been removed from the analysis in this paper. The reasons for these exclusions are usually strongly non-photometric observing conditions which alter the observed colour of stars or the use of filters which are too different for an accurate calibration without the inclusion of colour terms. 

\begin{figure*}
\centering
\includegraphics[angle=0,width=\columnwidth]{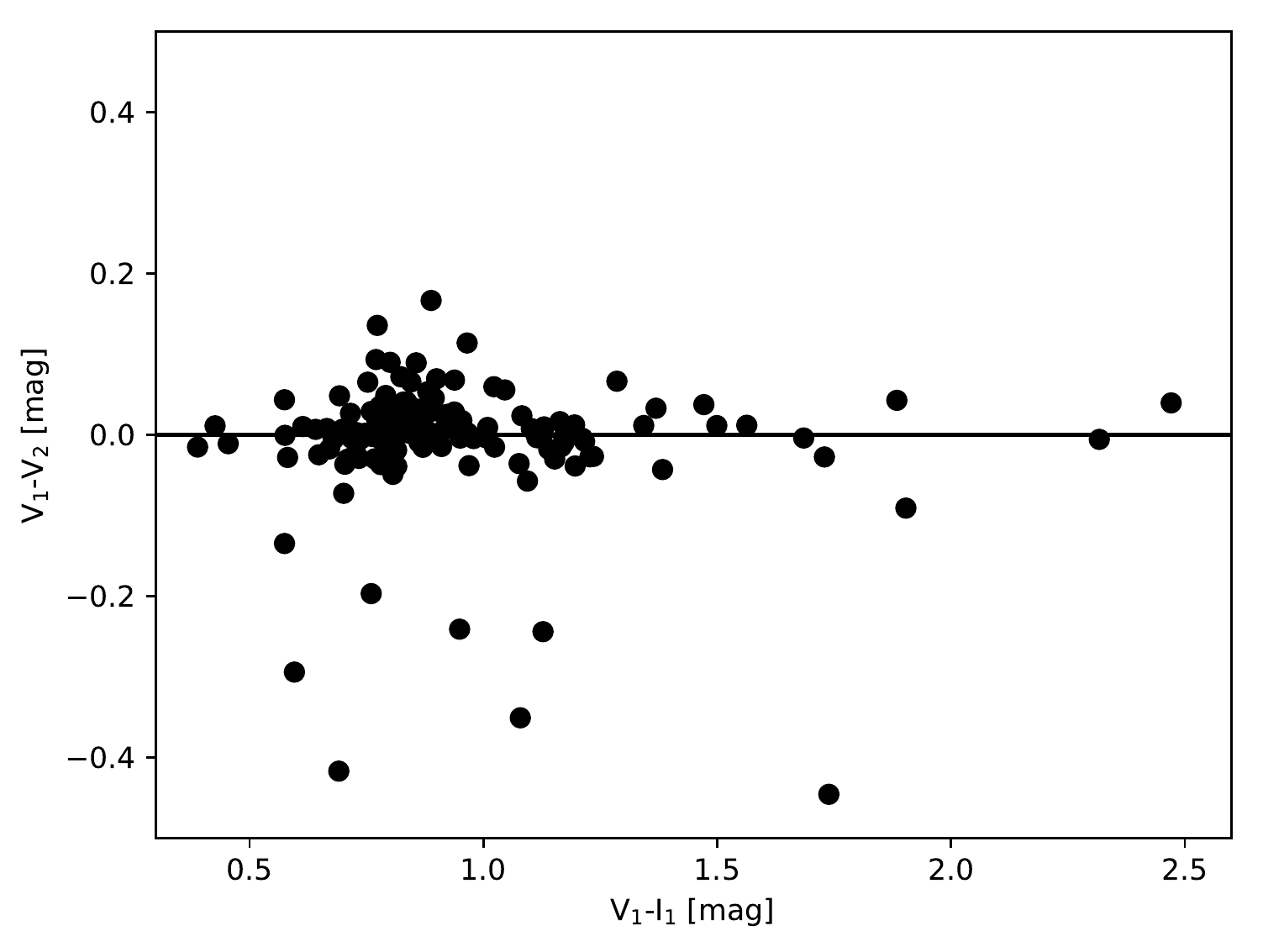} \hfill \includegraphics[angle=0,width=\columnwidth]{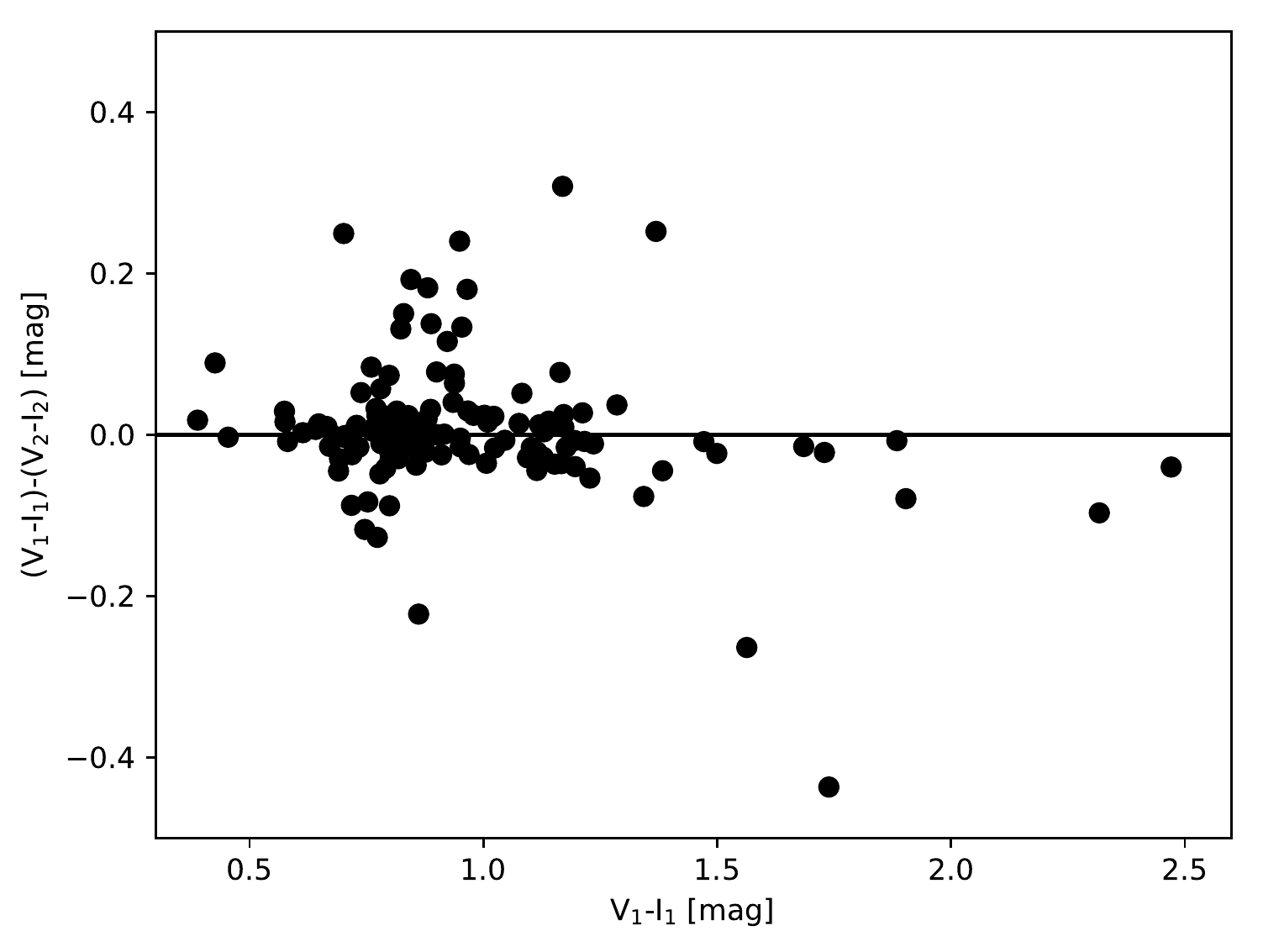}
\caption{Differences of calibrated instrumental V-Band magnitudes and V-I colours for one of our targets. An index of 1 indicates data taken with the Beacon Observatory, while an index of 2 indicates data taken with the LCO. There are no significant and systematic trends of these differences with the V-I colour of the stars. Note that the V and I data for each telescope has been taken on the same night, but there is an observing time difference of several weeks between the two datasets, thus the outlier data-points represent variable stars. \label{fig_example_calibration2}}
\end{figure*}

The absence of colour terms in the calibration will not influence the results obtained in this work. As indicated in Sect.\,\ref{hc_project}, the vast majority of data has been taken with a single instrument - the Beacon Observatory. All images with a large scatter in the calibration where removed from the analysis. For the investigation we only select stars with more than 0.2\,mag variability (see Sect.\,\ref{yso_selection}). To further ensure that no variability in the sources is induced by colour effects, all parameters determined for individual sources in the paper are based on observations in the standard V, R and I-Band filters only. We demonstrate in Fig.\,\ref{fig_example_calibration2} that there are no significant systematic trends of the calibrated instrumental V-band magnitudes and V-I colours between different telescope and filter combinations.

\begin{center}
\begin{table*}
\caption{Properties of star forming regions and clusters investigated in this work. In the Table we list the following: name; position in RA/DEC (J2000) as used in the HOYS-CAPS project; total number of potential YSOs detected in at least 15 images; number of variable stars selected (total number, number of YSO candidates and number of other variable stars with high Stetson index); fraction of YSO candidates that are variable; fraction of variable objects that are classified as CTTSs (f$^{\rm var}_{\rm CTTS}$) or transition/depletion disk objects (f$^{\rm var}_{\rm WTTS}$); distance in [pc]; age in [Myr]; The references for distances and ages for the different regions are indicated in the name column and are from: $^{(1)}$ \citet{2005AJ....130..188S}; $^{(2)}$ \citet{2002AJ....124.1585C}; $^{(3)}$ \citet{2008hsf1.book..308B}; $^{(4)}$ \citet{2008hsf1.book...36R}; $^{(5)}$ \citet{2009ApJ...697..787G}; $^{(6)}$ \citet{2008ApJ...680..495H}; $^{(7)}$ \citet{2013ApJ...775..138S}; $^{(8)}$ \citet{2008hsf1.book..928R}; $^{(9)}$ \citet{2008hsf1.book..966D}; $^{(10)}$ \citet{2014MNRAS.438.1848S}. Typically the distance uncertainties are 10\,--\,20\,\%, but they are probably much higher for IC\,5146. \label{table_clusters}}
\begin{tabular}{c|c|c|c|c|c|c|c|c|c|c|c}
\hline
Name & RA & DEC & N$_{\rm tot}$ & \multicolumn{3}{c|}{N$_{\rm var}$} & f$^{\rm YSO}_{\rm var}$ & f$^{\rm var}_{\rm CTTS}$ & f$^{\rm var}_{\rm WTTS}$ & d & age \\
 & \multicolumn{2}{c|}{(J2000)} & YSO & tot & YSO & var & [$\%$] & [$\%$] & [$\%$] & [pc] & [Myr] \\
\hline
IC\,1396\,A$^{(1,2)}$& 21 36 35 & +57 30 36 & 68 & 56 & 42 &14 & 62 & 34 & 66 & 900 & 1\,--\,5 \\
IC\,348$^{(3)}$      & 03 44 34 & +32 09 48 & 88 & 37 & 36 & 1 & 41 & 49 & 51 & 300 & 2\,--\,4 \\
IC\,5070$^{(4,5)}$   & 20 51 00 & +44 22 00 &152 &107 &105 & 2 & 69 & 87 & 13 & 600 & 3        \\
IC\,5146$^{(6)}$     & 21 53 29 & +47 16 01 &166 & 67 & 58 & 9 & 35 & 57 & 43 & 950 & 1\,--\,5 \\
NGC\,1333$^{(7)}$    & 03 29 02 & +31 20 54 & 30 & 19 & 17 & 2 & 57 & 74 & 26 & 300 & 1\,--\,3 \\
NGC2244$^{(8)}$      & 06 31 55 & +04 56 30 &544 & 65 & 54 &11 & 10 & 34 & 66 &1400\,--\,1700 & 2\,--\,3 \\
NGC2264$^{(9)}$      & 06 40 58 & +09 53 42 &124 & 79 & 73 & 6 & 59 & 67 & 33 & 760 & 1\,--\,5 \\
NGC\,7129$^{(10)}$   & 21 42 56 & +66 06 12 & 35 & 36 & 28 & 8 & 80 & 39 & 61 &1150 & 3        \\
\hline
\end{tabular}
\end{table*}
\end{center}

\section{Data analysis}\label{dataandanalysis}

\subsection{Selection of Young Clusters and their variable YSOs}\label{yso_selection}

For the analysis presented in this paper we selected the eight clusters/regions from the \hc\ target list which had the most observations available as of May 2017. These are: NGC\,7129, IC\,5070, NGC\,2264, IC\,1396\,A, IC\,348, NGC\,1333, IC\,5146 and NGC\,2244. We present the basic data for these clusters (positions, distances, ages, number variable stars selected etc.) in Table\,\ref{table_clusters}. 

In this paper we aim to generally characterise all variable YSOs in the above selected eight clusters. To achieve this we first cross-matched the optical photometry catalogues for our observed fields with catalogues of known and suspected members for each of the regions. In particular we used the following lists: \citet{2016ApJ...827...52L} for NGC\,1333 and IC\,348; \citet{2015AJ....149..200D} and \citet{2009A&A...507..227S} for NGC\,7129; \citet{2005AJ....130..188S} for IC\,1396\,A; \citet{2005AJ....129..829D} and \citet{2014AJ....147...82C} for NGC\,2264; \citet{2007ApJ...660.1532B} for NGC\,2244; \citet{2011ApJS..193...25R} for IC\,5070; and \citet{2008ApJ...680..495H} for IC\,5146. These catalogues are inhomogeneous in completeness, depth and selection criteria. In this work, they are only used to select variable young stars in these clusters which are detected in our optical data. 

We then investigated the light-curves of all these YSO candidate members in each region. As we are trying to analyse the statistical behaviour of variable young stars, only objects which fulfilled the following criteria where selected for further analysis: i) There are more than 15 nights of data with photometry in V, R and I (or equivalent filter) taken in the same night; ii) The difference in magnitude between the brightest and faintest V-band measurement has to be larger than 0.2\,mag (corresponding to about 5$\sigma$ for the brighter stars); iii) The Stetson index \citep{1996PASP..108..851S} has to be larger than one. These criteria ensure that our sample is not significantly contaminated by non-variable stars with large photometric errors. The numbers of targets selected by our criteria in each region are listed in Table\,\ref{table_clusters}. In each cluster/region we analysed some additional objects which are not in one of the catalogues but fulfill the above criteria for variability. Their numbers are indicated separately in Table\,\ref{table_clusters}.

\begin{figure*}
\centering
\includegraphics[angle=0,width=\textwidth]{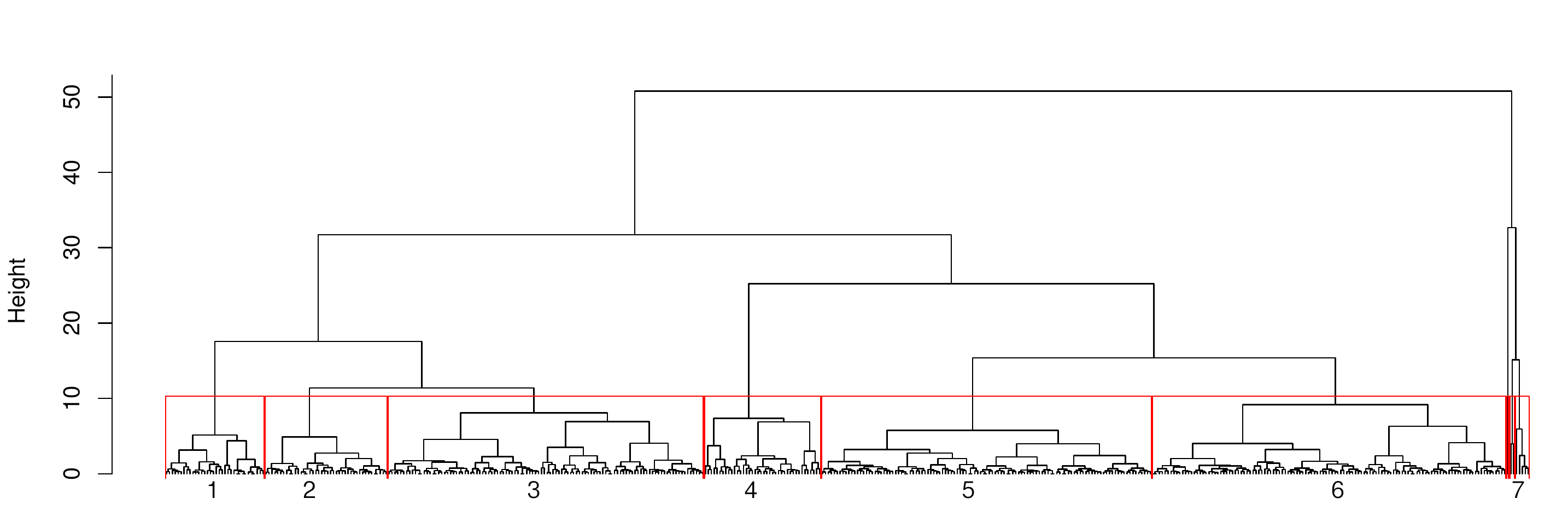} 
\caption{Resulting dendrogram of our hierarchical clustering analysis of the YSOs light-curves in our sample. The groups of YSOs identified are labeled at the bottom. All outliers on the right hand side of the dendrogram are combined into group\,7. The main Table\,\ref{yso_data_table} in the Appendix lists the group number for each individual YSO.\label{fig_dendrogram}}
\end{figure*}

\subsection{Determination of light-curve properties}

For each selected individual star we determined a number of light-curve properties from the available data. An example of some sets of light-curves can be found in Figs.\,\ref{V2492Cyg}, \ref{V350Cep}, and \ref{J2138}. These properties are the following:

\begin{itemize}

\item We fit a linear function (see Eq.\,\ref{eq_slope}) to the $N$ data-points in the V-I vs. V colour-magnitude diagram to determine the slope ($\alpha$) -- indicated as green lines in the right hand side panels of Figs.\,\ref{V2492Cyg}, \ref{V350Cep}, and \ref{J2138}. This has been done by minimising the sum over all perpendicular distances of the data-points to the line of best fit (see Eq.\,\ref{eq_errfct}) to ensure that for all extreme cases (colour independent magnitude changes and magnitude independent colour changes), the correct slope is determined. We determine the $rms$ value of all perpendicular distances to the line of best fit. We remove any outliers further than three times the $rms$ away from the line of best fit iteratively. These $rms$ values are almost exclusively larger than the photometric uncertainties of the individual data-points on both axes and thus only real outliers, such as in the case of V\,350\,Cep (Fig.\,\ref{V350Cep}) are removed. To ensure the procedure has worked for all objects, we manually inspected the graphs for all targets. For ease of presentation of the $\alpha$-values, we have converted them into degrees as otherwise some of the slopes have very large numerical values. Note that the photometric uncertainties of the data-points on both axes are correlated. This, however, has the effect of shifting data-points parallel to the $\alpha = 45^\circ$ directions, and has thus no significant influence on the determined values.

%Since there are some objects with basically no change in V-I colour but strong variability in V (vertical slope), this is not done by a linear regression, since this minimises the vertical distances of the data to the line of best fit. Instead we have minimised the distances of the data-points perpendicular to the line of best fit using a least-square optimisation method. This ensures the correct determination of the slope under all circumstances. Note that individual data-points that are more than 3$\sigma$ away from the line of best fit are removed iteratively during the fitting process. Due to the fact that the slopes can be almost vertical, or negative with very large values, we have converted them into degrees for the ease of presentation.

\begin{equation}\label{eq_slope}
V = V_0 + \alpha (V-I)
\end{equation}

\begin{equation}\label{eq_errfct}
\sum\limits^N_{i=1} \frac{\left| V_0 (V-I) + \alpha - V \right|}{\sqrt{V_0^2 + 1}}
\end{equation}

\item After the calculation of the line of best fit and removal of outlying data-points, we determine the scatter ($rms$) of the data-points in the V-I vs V colour-magnitude diagram perpendicular to the line of best fit.

\item Following \citet{2014AJ....147...82C} we determine the asymmetry metric ($M$) for the V-Band light-curve using $M = \left(  \left< d_{10\,\%} \right> - d_{\rm med}  \right) / \sigma_d$, where $\left< d_{10\,\%} \right>$ is the mean of all magnitudes in the top and bottom ten percent of the V-Band light-curve, $d_{\rm med}$ the median of all magnitudes and $\sigma_d$ the overall rms of the V-Band data-points from the mean. 

\item Following \citet{1996PASP..108..851S} we determine the Stetson index for the V-Band light-curve.

\item We determine a cumulative distribution function (CDF) for the V-Band magnitudes after subtracting the median magnitude from each data-point. 

\end{itemize}

We then use a two-sided, two sample Kolmogorov-Smirnov (KS) test to compare the CDFs of all objects against each other pairwise and record the KS statistics (D$_{\rm KS}$) and resulting $p$-value. This value indicates the probability that the two samples (V-Band CDFs) are drawn from the same parent distribution. Hence, large $p$-values indicate pairs of stars with very similar distribution of V-Band magnitudes and vice versa. 

\subsection{YSO near and mid-IR SEDs}

We have cross matched the objects investigated here with the WISE All-sky catalogue \citep{2013yCat.2328....0C} and the 2MASS catalogue \citep{2006AJ....131.1163S}. For some of our target regions there are {\it Spitzer} observations which are deeper than the WISE data. However, they are not available for all fields to a homogeneous depth, which we have for our optical data. Since more than 90\,\% of our objects have a WISE match in all four filters, we do not include the {\it Spitzer} data in our analysis. The WISE data has been used to determine the slope of the spectral energy distribution ($\alpha_{\rm SED}$) between 3.4\,$\mu$m and 22\,$\mu$m. Following \citet{2013Ap&SS.344..175M} we use:

\begin{equation}
\alpha_{\rm SED} = 0.36\cdot(w1-w2)+0.58\cdot(w2-w3)+0.41\cdot(w3-w4)-2.90
\end{equation}

Traditionally, objects with positive slopes $\alpha_{\rm SED}$ are classified as protostars, while negative slopes between $-2$ and zero indicate classical T-Tauri stars (CTTSs). \citet{2014ApJ...791..131K} have presented a colour selection scheme in WISE to differentiate Class\,I protostars and Class\,II CTTSs based not just on $\alpha_{\rm SED}$ but considering all WISE colours (excluding the W4 magnitudes). There is no specific WISE colour selection for Weak-Line T-Tauri Stars (WTTS) as their colours overlap too much with normal stars as well as other objects such as AGB stars.

\subsection{Hierarchical clustering of YSO light-curve properties} 

We aim to identify objects with similar light-curve properties automatically via a hierarchical clustering method. Here we briefly describe the methods involved. A more detailed discussion of them can be found in Campbell-White et al. (2018, subm.).

As a first step we establish a dissimilarity matrix which contains the pairwise distances ($d_{\rm (i,j)}$) between each pair of YSOs (i,j) for each of the properties. From the KS-test we already have $d^{\rm KS}_{\rm (i,j)}$ which is identical to the KS statistic D$_{\rm KS}$. We also determine these distances for each of the other properties, i.e. $d^{\alpha}_{\rm (i,j)} = \left| \alpha_{\rm i} - \alpha_{\rm j} \right|$ for the slopes in the V-I vs V colour magnitude diagram; $d^{rms}_{\rm (i,j)} = \left| rms_{\rm i} - rms_{\rm j} \right|$ for the scatter of the data-points from the line of best fit in the same diagram, and $d^{M}_{\rm (i,j)} = \left| M_{\rm i}- M_{\rm j} \right|$ for the asymmetry metric of the V-Band light-curves.

To account for the different ranges in these four parameters, all the individual distances are normalised by the mean of all pairwise distances for the respective parameter. The final pairwise dissimilarity between two stars, considering all of the parameters is then calculated as:

\begin{equation}
d_{\rm (i,j)} = \sqrt{  
\left( \frac{d^{\rm KS}_{\rm (i,j)}}{\overline{d^{\rm KS}}} \right)^2  +  
\left( \frac{d^{\alpha}_{\rm (i,j)}}{\overline{d^{\alpha}}} \right)^2  +  
\left( \frac{d^{rms}_{\rm (i,j)}}{\overline{d^{rms}}} \right)^2 +  
\left( \frac{d^{M}_{\rm (i,j)}}{\overline{d^{M}}} \right)^2 }
\end{equation}

An Euclidean distance matrix is then calculated from the $d_{\rm (i,j)}$-values (see Campbell-White et al. (2018, subm.) for details) and hierarchical clustering is performed on it to identify groups of YSOs with similar properties. Note that we will use 'groups' throughout the paper when referring to these statistical associations, rather than 'cluster/region' which refers to the investigated young clusters from which the YSOs are selected. 

Groups are created from hierarchical clustering by considering, in the first instance, the distance between pairs of objects. Pairs with the smallest distances are joined, forming the first groups. Recursive merging then takes place, where at each stage, new inter-group distances are determined. The manner in which this happens depends on which hierarchical method is used. We have used Ward's agglomerative method to form groups \citep{murtagh2014ward}, which seeks to minimise the distance between the centres of groups in Euclidean space, and has been shown to outperform other hierarchical methods \citep{ferreira2009comparison}. Groups are iteratively formed until the final two groups merge. This process is represented by a dendrogram. Figure\,\ref{fig_dendrogram} shows all individual YSOs along the bottom, and groups are seen as the horizontal merges. The height on the dendrogram corresponds to the distance at which a merge was made, for pairs of YSOs, this was the Euclidean distance we have described, for all higher level groups, the height was taken from the value obtained with Ward's method.

\begin{figure*}
\centering
\includegraphics[angle=0,width=\columnwidth]{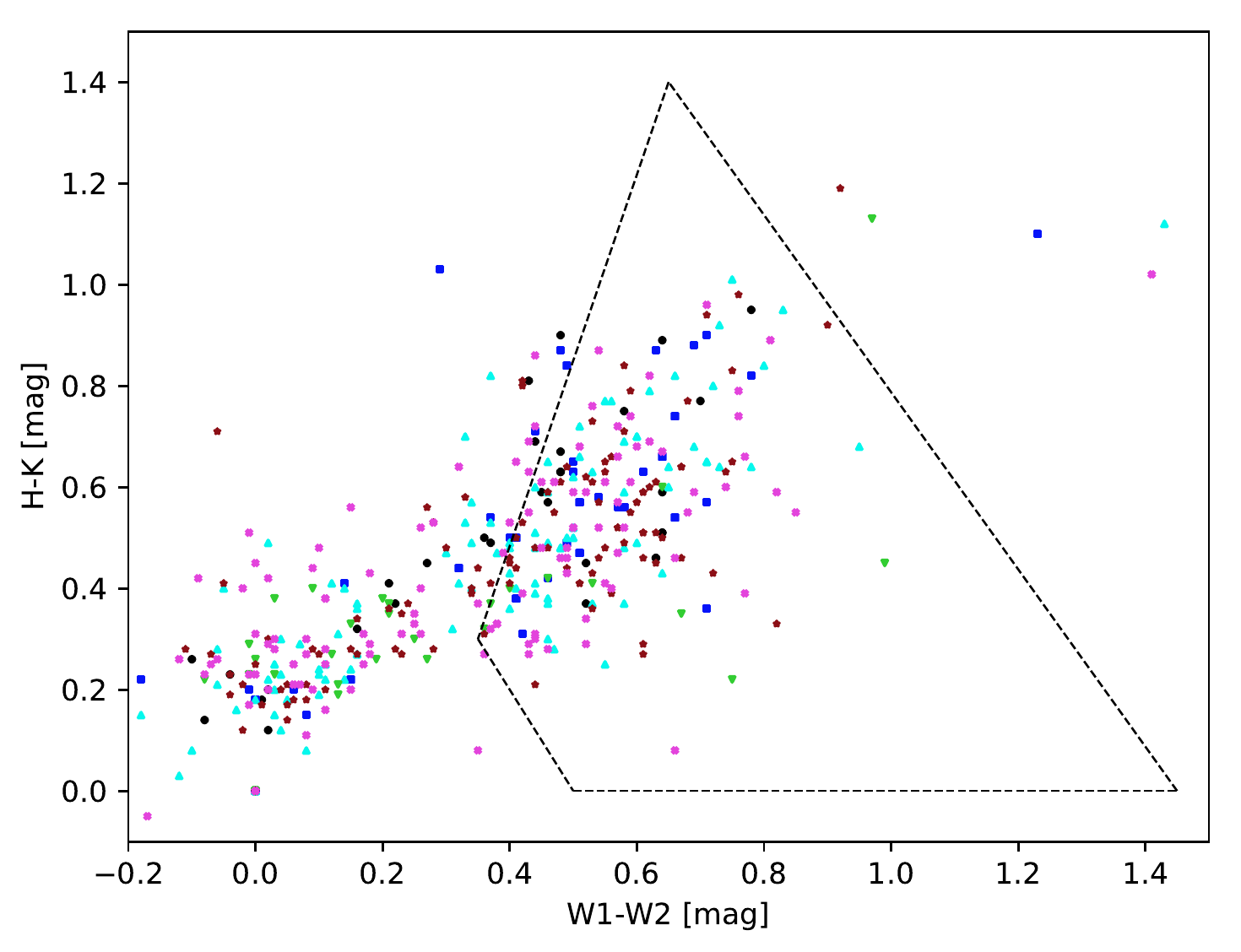} \hfill
\includegraphics[angle=0,width=\columnwidth]{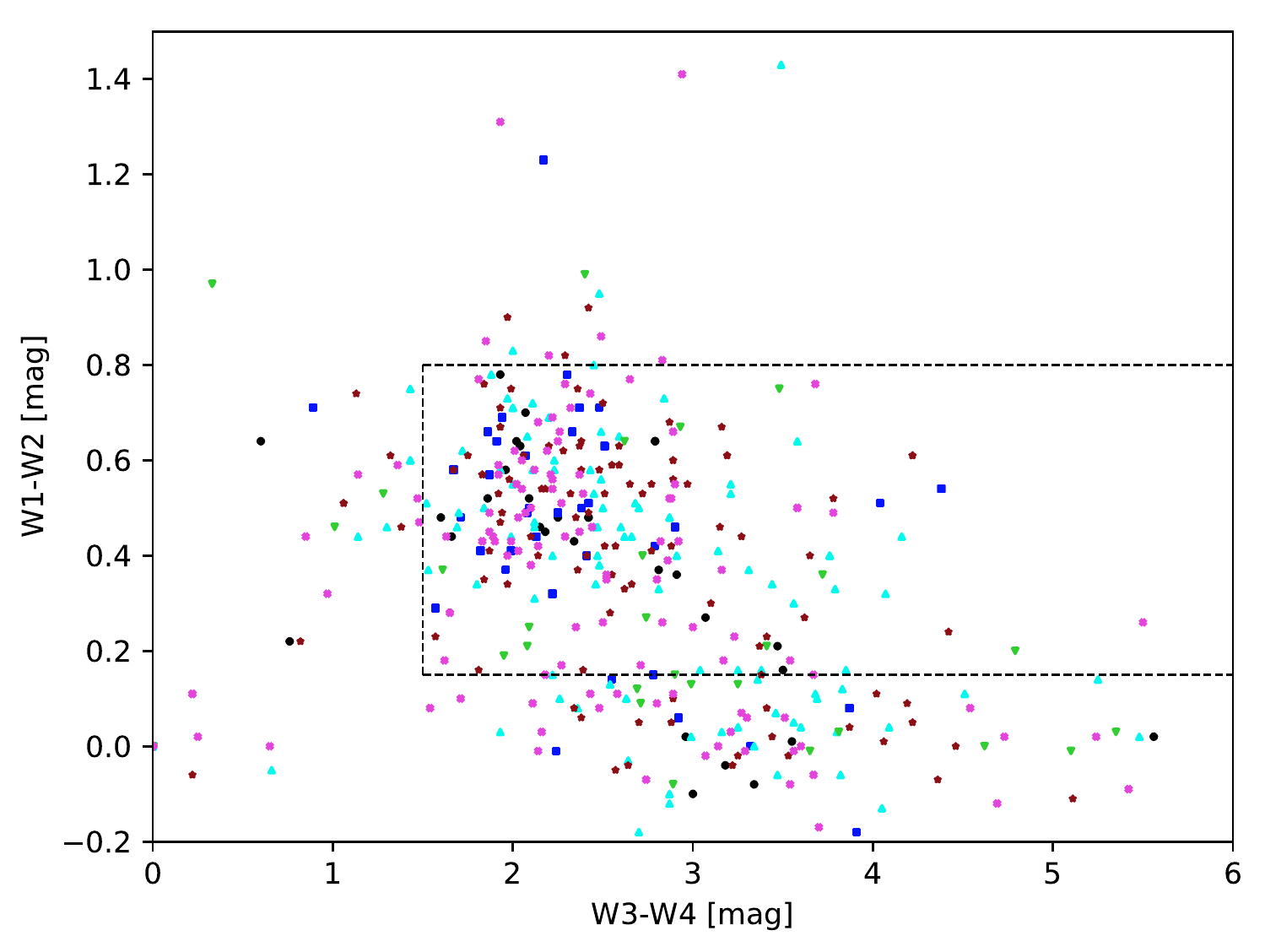} 
\caption{NIR/MIR colour-colour diagrams of the variable objects investigated. {\bf Left:} WISE W1-W2 vs H-K 2MASS colour-colour plot. The dashed black lines enclose the region of CTTSs from \citet{2014ApJ...791..131K}. Class\,I protostars would be situated towards the top-right part of the diagram. {\bf Right:} WISE W3-W4 vs W1-W2 diagram. The dashed black lines enclose the region for transition disks from \citet{2014ApJ...791..131K}. The two distinct groups of sources, separated by W1-W2\,=\,0.3\,mag, can be clearly seen in both panels. The colours and symbols indicate membership in each of the groups identified with the hierarchical clustering: G\,1 -- black circles; G\,2 -- dark blue squares; G\,3 -- light blue triangles up; G\,4 -- green triangles down; G\,5 -- brown stars; G\,6 -- pink x's; All outliers in G\,7 are not shown. \label{fig_yso_ccds}}
\end{figure*}

\section{Results and Discussion}\label{results}

\subsection{Selection of variable YSOs}

Our selection criteria for variable YSOs in the eight investigated regions have identified 466 objects which are included in the analysis in this paper. Of these, 413 (89\,\%) are included in one of the catalogues of potential cluster members, the remaining 53 objects (11\,\%) are included solely due to their variability. Hence, the sample of variables investigated will only contain a very small fraction of potentially non-YSO variables such as background giant stars. In Table\,\ref{table_clusters} we detail the number of investigated variable stars in each of the clusters, split by their mode of selection. This table can be used as a compendium of variable YSOs in future studies, together with the main Table\,\ref{yso_data_table} in the Appendix where all individual stars and their properties are listed.

We estimate the fraction of variable stars (f$^{\rm YSO}_{\rm var}$) amongst the detected YSOs. These fractions are also listed in Table\,\ref{table_clusters}. The mean fraction of variable stars is 52\,\% with a rms variation of 22\,\% between the individual regions. The highest fraction of variables (with 80\,\%) occurs in NGC\,7129. In NGC\,2244 the fraction is the lowest with only 10\,\% of the detected stars classified as variable. These numbers are only included here for completeness reasons, since (as indicated above) the strong variations in them are most likely a reflection of the our selection method and biases than that they are intrinsic to the properties of the regions (such as the age). Variability is expected to decline with the age of the region, and our ability to detect it will depend on the distance and extinction.

%\as{The following paragraph could be removed, because it doesn't go anywhere.}
%We further aim to characterise the typical evolutionary status of the variable YSO population in our sample. Using the determined slope of the spectral energy distribution in the WISE bands ($\alpha_{\rm SED}$) and the characterisation as protostars or CTTSs from \citet{2014ApJ...791..131K} does not lead to any useful results. Only six objects are classified as protostars -- one each in IC\,1396\,A and NGC\,7129, as well as four in IC\,5070. However, there are many more objects with a positive $\alpha_{\rm SED}$ value. This is caused by the fact that most protostars will be extremely faint in the optical and are thus not detected in our data. Furthermore, the variability of the sources will cause some mis-classifications.

The colour-colour diagrams shown in Fig.\,\ref{fig_yso_ccds} reveal that there are two distinct groups of objects in our sample of variable YSOs. The left panel of Fig.\,\ref{fig_yso_ccds} shows the  W1-W2 vs H-K diagram. It indicates the typical colours of CTTS from \citet{2014ApJ...791..131K} and a large number of objects fall in or near the CTTS region. Note that protostars would be situated towards redder colours along both axes. A second group of objects with almost zero W1-W2 colour is also evident. The same two groups of objects can be identified in the right panel of Fig.\,\ref{fig_yso_ccds} which shows the W3-W4 vs W1-W2 colour-colour diagram. One finds that the objects with low W1-W2 colours, falling below the marked box, show in part large W3-W4 values, thus indicating the presence of some cold disk material. These objects are thus most likely either transition disks with inner gaps or depleted disks. 

Thus, for the purpose of this paper we characterise all sources with W1-W2\,$>$\,0.3\,mag as CTTS and all other sources as transition/depleted disks (referred to as WTTS, hereafter). Note that this criterion is only applicable as the vast majority of our sources are selected as known YSOs which are also variable, hence there is almost no contamination of non-YSOs in the sample. The fraction of CTTSs amongst the variables has been listed as f$^{\rm var}_{\rm CTTS}$ in Table\,\,\ref{table_clusters}. The mean CTTS fraction in the full sample is 64\,\% (271 objects). The highest value in an individual region (87\,\%) is found in IC\,5070 and the lowest (34\,\%) in IC\,1396\,A and NGC\,2244. There are no apparent trends between these fractions and the age of the cluster/region they are in. 

\begin{center}
\begin{table*}
\caption{Table with data of number/fraction of YSOs per region/cluster and group in dendrogram. We list the number ($N$) of objects in each of the groups from each young region/cluster. The first number in the brackets indicates what fraction $N$ represents in the region/cluster expressed in percent. The second number indicates what fraction $N$ represents in the group. For example for IC\,5070 in G2 we find 13 (12,31), which means there are 13 objects in the region IC\,5070 which are part of G\,2 and these sources represent 12\,\% of all the objects in IC\,5070 and 31\,\% of all the objects in G\,2. The numbers in brackets in the 'Total' row and column indicate the fraction of the total number of source in the sample. Note that due to rounding errors the fractions do not necessarily add up to 100\,\%. \label{table_groups} }
\begin{tabular}{c|c|c|c|c|c|c|c|c}
\hline
Cluster & G1 & G2 & G3 & G4 & G5 & G6 & G7 & Total \\
\hline
IC\,1396\,A & 4 ( 7,12) & 7 (13,17) &17 (30,16) & 4 ( 7,10) &11 (20,10) &12 (21,10) & 1 ( 2,13) & 56 (12) \\
IC\,348     & 0 ( 0, 0) & 2 ( 5, 5) & 6 (16, 6) & 7 (19,18) & 6 (16, 5) &14 (38,12) & 2 ( 5,25) & 37 ( 8) \\
IC\,5070    & 4 ( 4,12) &13 (12,31) &20 (19,19) & 5 ( 5,13) &40 (37,35) &25 (23,21) & 0 ( 0, 0) &107 (23) \\
IC\,5146    &10 (15,29) & 5 ( 7,12) &16 (24,15) & 1 ( 1, 3) &18 (27,16) &17 (25,14) & 0 ( 0, 0) & 67 (14) \\
NGC\,1333   & 3 (16, 9) & 2 (11, 5) & 4 (21, 4) & 0 ( 0, 0) & 2 (11, 2) & 7 (37, 6) & 1 ( 5,13) & 19 ( 4) \\
NGC\,2244   & 5 ( 8,15) & 4 ( 6,10) &18 (28,17) & 6 ( 9,15) &15 (23,13) &17 (26,14) & 0 ( 0, 0) & 65 (14) \\
NGC\,2264   & 8 (10,24) & 9 (11,21) &20 (25,19) & 8 (10,20) &14 (18,12) &20 (25,17) & 0 ( 0, 0) & 79 (17) \\
NGC\,7129   & 0 ( 0, 0) & 0 ( 0, 0) & 7 (19, 6) & 9 (25,23) & 7 (19, 6) & 9 (25, 7) & 4 (11,50) & 36 ( 8) \\
\hline
Total       & 34 ( 7)   & 42 ( 9)   &108 (23)   & 40 ( 9)   &113 (24)   &121 (26)   & 8 ( 2)    &466 \\
\hline
\end{tabular}
\end{table*}
\end{center}

\subsection{YSO groups from the Hierarchical Clustering}\label{res_clustering}

The hierarchical clustering results are presented in the dendrogram in Fig.\,\ref{fig_dendrogram}. Usually the dendrogram is cut at a specific, albeit arbitrary, height to select groups of objects. We have chosen a value of 10 as the height at which the cut is made. This results in 6 clear groups (referred to as G hereafter) and some outliers. In Table\,\ref{yso_data_table} we list for each individual YSO which group it belongs to. One way of justifying the choice of the cut value is to read the dendrogram from the top down and to investigate if the identified groups share any physical properties.

\begin{figure}
\centering
\includegraphics[angle=0,width=\columnwidth]{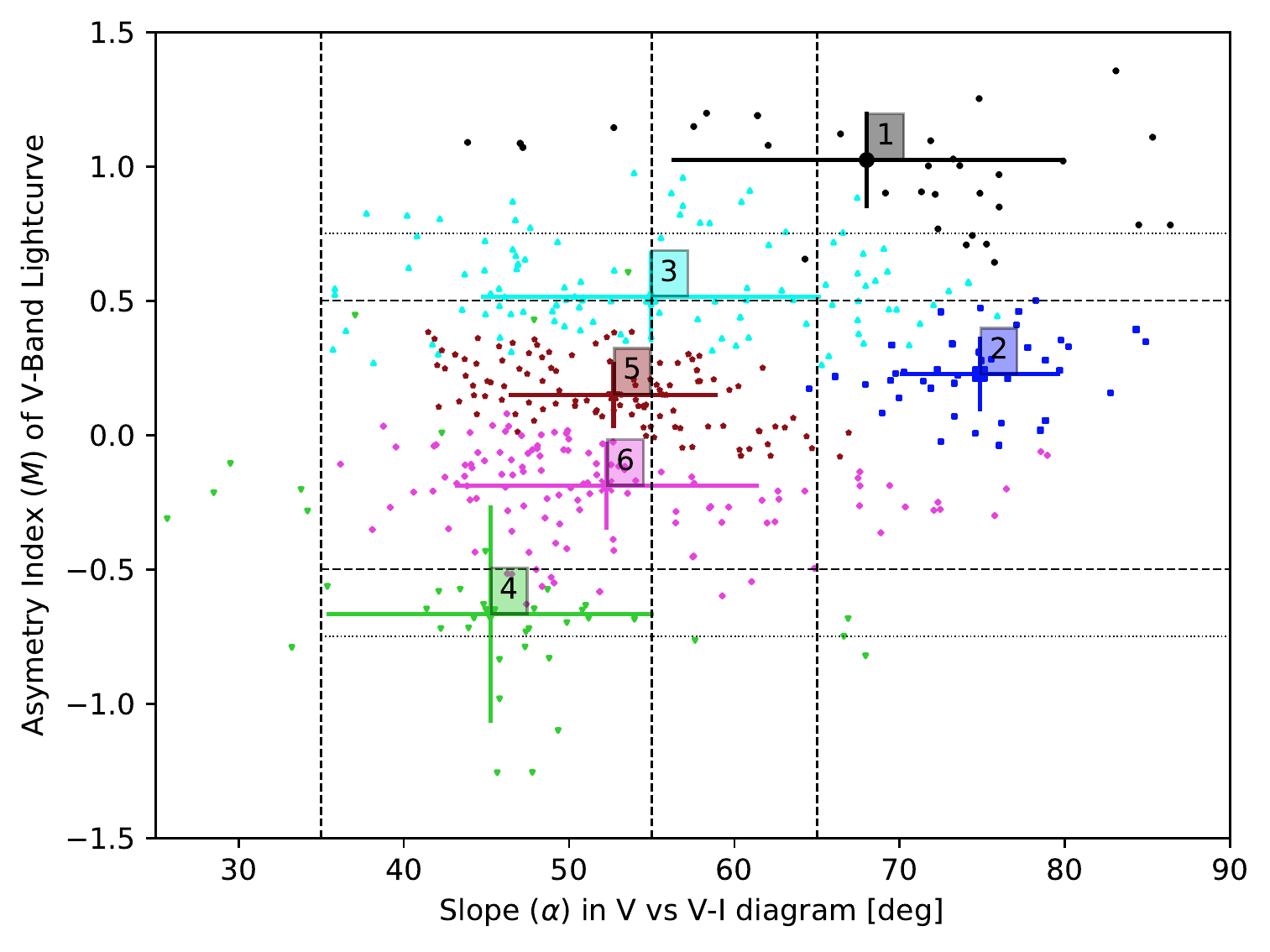} 
\caption{Figure showing the slope $\alpha$ in the V vs V-I diagram and the asymmetry index $M$ for the YSOs in our sample. The larger symbols and error bars indicate the mean and $rms$ of all stars in the different groups. The group number is also indicated. All outliers (summarised in G\,7) are not shown as they are partly outside the parameter space of the plot (e.g. at negative slope values). The colours and symbols are the same as in Fig.\,\ref{fig_yso_ccds}. The dashed horizontal lines separate the dippers (top) from the symmetric light-curves (middle) and the bursters (bottom). The thin dashed horizontal lines separate the most extreme bursters and dippers. The dashed vertical lines indicate the three regions for $\alpha$ discussed in the text. \label{fig_slope_vs_asym}}
\end{figure}

The first branch that separates from the majority of objects is seen on the right hand side and contains 8 objects. These objects have basically no common properties with all of the remaining sources and are hence placed in G\,7 (labeled in Fig.\,\ref{fig_dendrogram} together with all the other groups). The next separation in the dendrogram splits Gs\,1, 2, 3 from Gs\,4, 5, 6. As will be discussed later and indicated in Fig.\,\ref{fig_slope_vs_asym}, this splits the light-curves with dips from the light-curves which are symmetric or have outbursts. 

The next split in the objects with dipper light-curve separates G\,1 from the rest. According to Fig.\,\ref{fig_slope_vs_asym}, G\,1 contains the most extreme dippers. The final split into G\,2 and G\,3 separates the objects with almost colour-independent dips (G\,3) from the remaining dippers (G\,2). Similarly, the first split of the burster light-curves separates the most extreme bursters (G\,4) from the rest, which is then split into less extreme burster and symmetric light-curves (Gs\,6, 5).

\subsection{$\alpha$-values and $M$-values}

In Fig.\,\ref{fig_slope_vs_asym} we have plotted the positions of all variable objects (except the outliers in G\,7) in the $\alpha$\,--\,$M$ plane of the parameter space. Objects from the different groups are indicated by different colours and symbols. We also indicate the average values and $rms$ of all objects in each group. As one can see from this diagram and the discussion above, the hierarchical clustering seems to have separated the groups most clearly in this part of the parameter space. There is only minimal overlap between the groups with the exception of some scattered objects in G\,4. The groups are (with the exception of G\,2), separated by the asymmetry parameter $M$. Furthermore, the parameter space is not filled homogeneously. There are almost no sources with $\alpha$-values lower than 40\dg\ and objects with negative $M$-values and high $\alpha$-values are extremely rare.

The value of the $\alpha$ parameter indicates the amount of colour change during brightness change. Objects with large values (close to 90\dg) basically do not change their colour at all. These objects can for example be eclipsing binaries where both components have similar colours. Changes in brightness due to variability in the amount of absorbing and scattering material along the line of sight (e.g. from the accretion disk) generally changes the colour in a determinable way. If the disk material is made from normal ISM dust then the $\alpha$-value should be between 60\dg\ ($R_V = 3.1$) and 66\dg\ ($R_V = 5.0$) if one uses a standard reddening law \citep{1990ARA&A..28...37M}. Higher values of $\alpha$ are possible if the obscuring material consists of larger dust grains than in the normal ISM or the objects undergo eclipse events by optically thick material. 

Light-curves with $\alpha < 55$\dg\ are thus not consistent with variability due to changes in extinction alone. Objects that show $\alpha \approx 45$\dg\ change their brightness in the V-Band but not at a detectable level in the I-Band. This can in principle be caused by non-variable red sources which are at the detection limit in V but detected at high signal-to-noise in the I-Band. However, our selection criteria for the variable sources will have removed these reliably. Alternatively, these objects can be interpreted as sources with large fluctuations at visual wavelengths and small, undetected variations in the infrared. Hence, the most likely cause for the variability in objects with $\alpha < 55$\dg\ is changes in the accretion rate or the properties of large hot/cold spots. Thus, based on the $\alpha$-value we can classify the sources in the following three categories: 

\begin{itemize}
\item 65\dg$ \le \alpha \le 90$\dg: Light-curves consistent with variable extinction due to larger grains or eclipses by optically thick material
\item 55\dg$ \le \alpha \le 65$\dg: Light-curves consistent with variable extinction due to normal ISM dust grains
\item 35\dg$ \le \alpha \le 55$\dg: Light-curves not consistent with variable extinction, thus most likely caused by changes in accretion rates or hot/cold surface spots
\end{itemize}

Naturally there will be sources whose properties are a mix of more than one of these physical reasons for variability. Hence these categories will only be indicative of what the main reason for the variability in a group of objects might be if it falls within one of the specified regions for the $\alpha$-values. In total there are 252 objects whose $\alpha$-value is in agreement of their variability being dominated by accretion rate changes or spots. In the group that is dominated by variable ISM extinction there are 94 sources, and in the group of objects which are most likely due to variable extinction by larger grains or eclipses we have 108 sources. The remaining objects fall outside this part of the parameter space. 

We can also split the sample of sources according to the asymmetry metric $M$. As detailed in \citet{2014AJ....147...82C} objects with negative values are bursters, i.e. objects that show short duration increases in brightness compared to their normal magnitudes. On the other hand, objects with positive values for $M$ are dippers, which have short duration decreases in brightness compared to normal. \citet{2014AJ....147...82C} use $\pm 0.25$ as thresholds for $M$ to identify bursters and dippers. Here we employ a slightly more conservative value and will identify an object as burster if $M < -0.5$, while objects with $M > 0.5$ are considered dippers. Objects with $M$-values in-between these borders are considered to have a symmetric behaviour. In total our sample contains 41 bursters, 101 dippers and 324 symmetric variables. If we use an even more conservative threshold of $M \pm 0.75$ to select the most extreme bursters and dippers we find 10 and 51, respectively.

From this discussion and from Fig.\,\ref{fig_slope_vs_asym} it is clear that our objects do not fill the parameter space homogeneously. In particular there is not a single object whose properties are in agreement with variability due to eclipse events ($\alpha$ near 90\dg), which at the same time belongs into the burster group. This is understandable, since almost all eclipsing binary objects spend most of their time outside the eclipse state and eclipsing binaries that also show outbursts might be characterised as having a 'symmetric' light-curve. 

As already briefly discussed in Sect.\,\ref{res_clustering} the groups of YSOs generated by the hierarchical clustering, do not have a simple one-to-one matching with any one of the physical characteristics, but there are clear trends. Most of the objects in G\,4 fall into the burster category and most of them also are in agreement with having variable accretion rates or spots. Indeed more than 50\,\% of the G\,4 members are in the accretion burst/spot category. Similarly, members in G\,1 are almost exclusively in agreement with dipper light-curves, most of them in the extreme category. Most of the light-curves in G\,1 are either in agreement with variable extinction or eclipses, only four sources fall into the variable accretion rate/spot category. All the sources in G\,2 fall into the eclipse category, and all of the light-curves are symmetric. The objects sorted into the other groups (G\,3, 5, 6) are all a mix of accretion/spot and extinction variability, where G\,3 contains objects with slightly dipping light-curves. G\,5/G\,6 contain symmetric light-curves with slightly positive $M$-values in G\,5 and negative ones in G\,6.

\begin{figure}
\centering
\includegraphics[angle=0,width=\columnwidth]{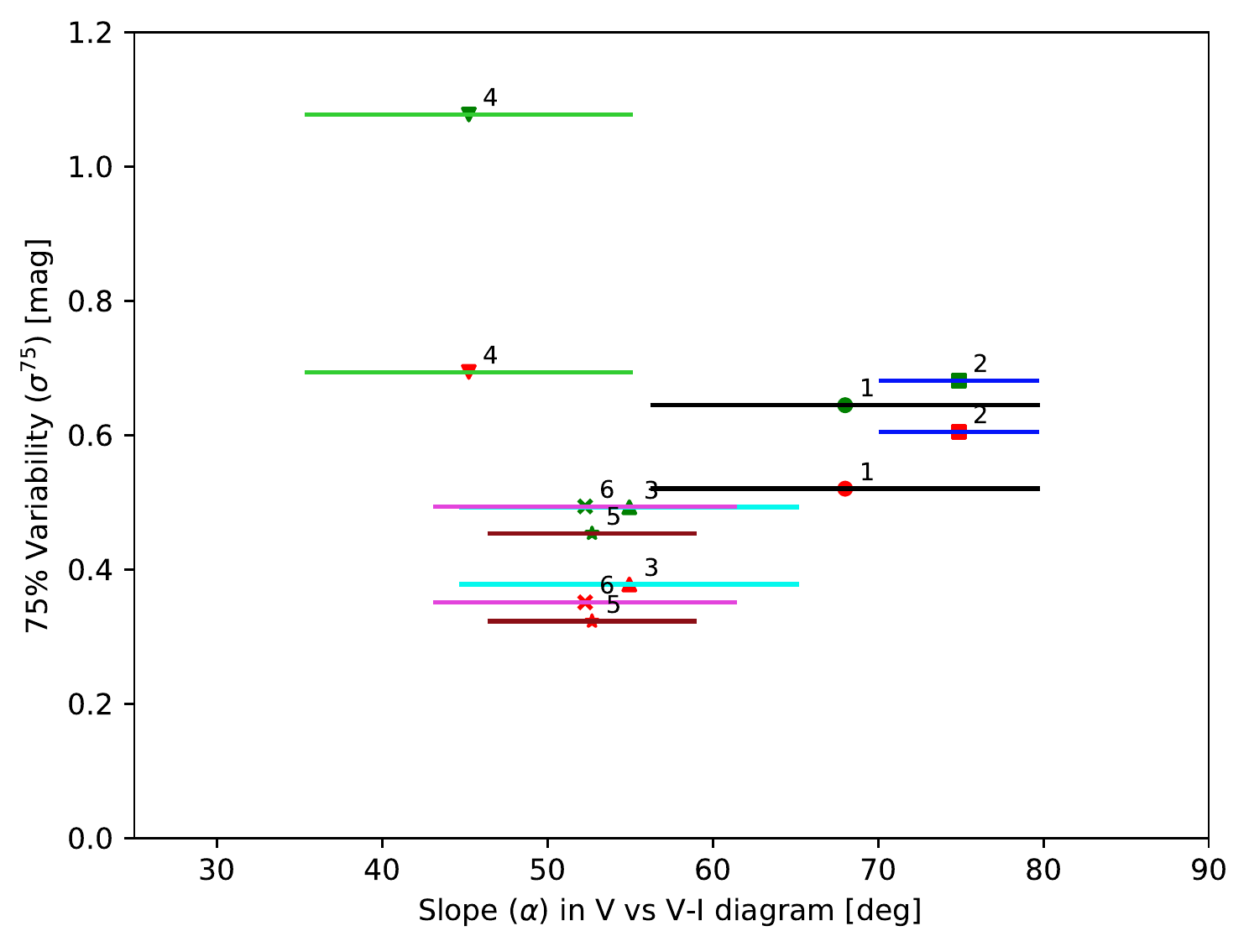} 
\caption{Figures showing the 75\,\% variability ($\sigma^{75}$) in V and R for the groups identified with the hierarchical clustering. The green symbols indicate $\sigma^{75}_V$ and the red symbols $\sigma^{75}_R$. The symbol types, numbers and colour of the lines are the same as in Fig.\,\ref{fig_yso_ccds}. \label{fig_variability}}
\end{figure}

\subsection{Variability}

We determine the typical variability in the V-Band and R-Band for groups of YSOs identified in the hierarchical clustering. This utilises the total CDF for all magnitudes determined as the average of all individual CDFs of all stars in each group. We then determine the magnitude range around the median value within which 75\,\% of the brightness measurements are situated and refer to this from here on as 75\,\% variability or $\sigma^{75}$.

In Fig.\,\ref{fig_variability} we show $\sigma^{75}$ of all groups in the V and R-Band plotted against the average $\alpha$-value. In all cases, $\sigma^{75}_{V}$ is larger than $\sigma^{75}_{R}$ by at least about 0.1\,mag. The lowest variability (about 0.5\,mag in V and 0.4\,mag in R) is measured for Gs\,3, 5, 6 which represent the symmetric light-curves. Slightly higher variability (0.7\,mag in V and 0.6\,mag in R) is found for Gs\,1, 2. These are all objects whose light-curves can be characterised as dippers and all of them are also in agreement with variable extinction or eclipse behaviour. 

The most variable sources (about 1.1\,mag in V and 0.7\,mag in R) are the objects in G\,4, i.e. the objects most likely in agreement with outbursts due to variable accretion rates (the magnitudes of the variability in this group exclude spots as causes in most cases). Objects in this group also show the largest difference $\sigma^{75}_{V}-\sigma^{75}_{R}$. While for all other groups there is only a 0.1\,mag difference between the V and R filters, for G\,4 the difference is 0.4\,mag. This can only in part be explained by the low $\alpha$-values of the sources, as especially Gs\,5, 6 have similar values, but the difference in variability from R and V is much smaller, both in absolute and relative terms.

\subsection{YSO properties in Clusters/SF Regions}

In Table\,\ref{table_groups} we have indicated for each group from the hierarchical clustering the number of members and how they are distributed amongst the investigated young clusters and SF regions. We find that there are three groups (Gs\,2, 5, 6) which roughly contain one quarter of all the sources each. The remaining three groups (Gs\,1, 2, 4) each roughly contain 10\,\% of all the objects. The fraction of sources considered outliers (summarised in G\,7) is extremely small with less than 2\,\% but this group naturally contains the most unusual objects in the sample. 

We have checked if any cluster or SF region has an abnormally high/low representation of objects in a particular group. There are a two notable cases: i) A quarter of all objects from NGC\,7129  are in G\,4, while this group only contains 9\,\% of all objects; ii) Half of all objects in G\,7 are from NGC\,7129, while this region/cluster only represents 8\,\% of all objects. Thus, NGC\,7129 seems to be unusual in that it contains a larger fraction of YSOs whose light-curves can be characterised as bursters caused by accretion rate variations. The cluster also seems to contain a high fraction of objects with unusual light-curves. However, small number statistics as well as selection biases could be responsible for these abnormalities.

An investigation into potential differences in the light-curve properties of CTTSs and WTTSs in our sample has not led to any significant results. There are hints that the fraction of objects whose light-curves can be interpreted as being caused by occultations by material made up of larger dust grains is increased amongst the WTTS population. This is not statistically significant, however it does warrant a more detailed investigation in future.

\begin{figure}
\includegraphics[angle=0,width=\columnwidth]{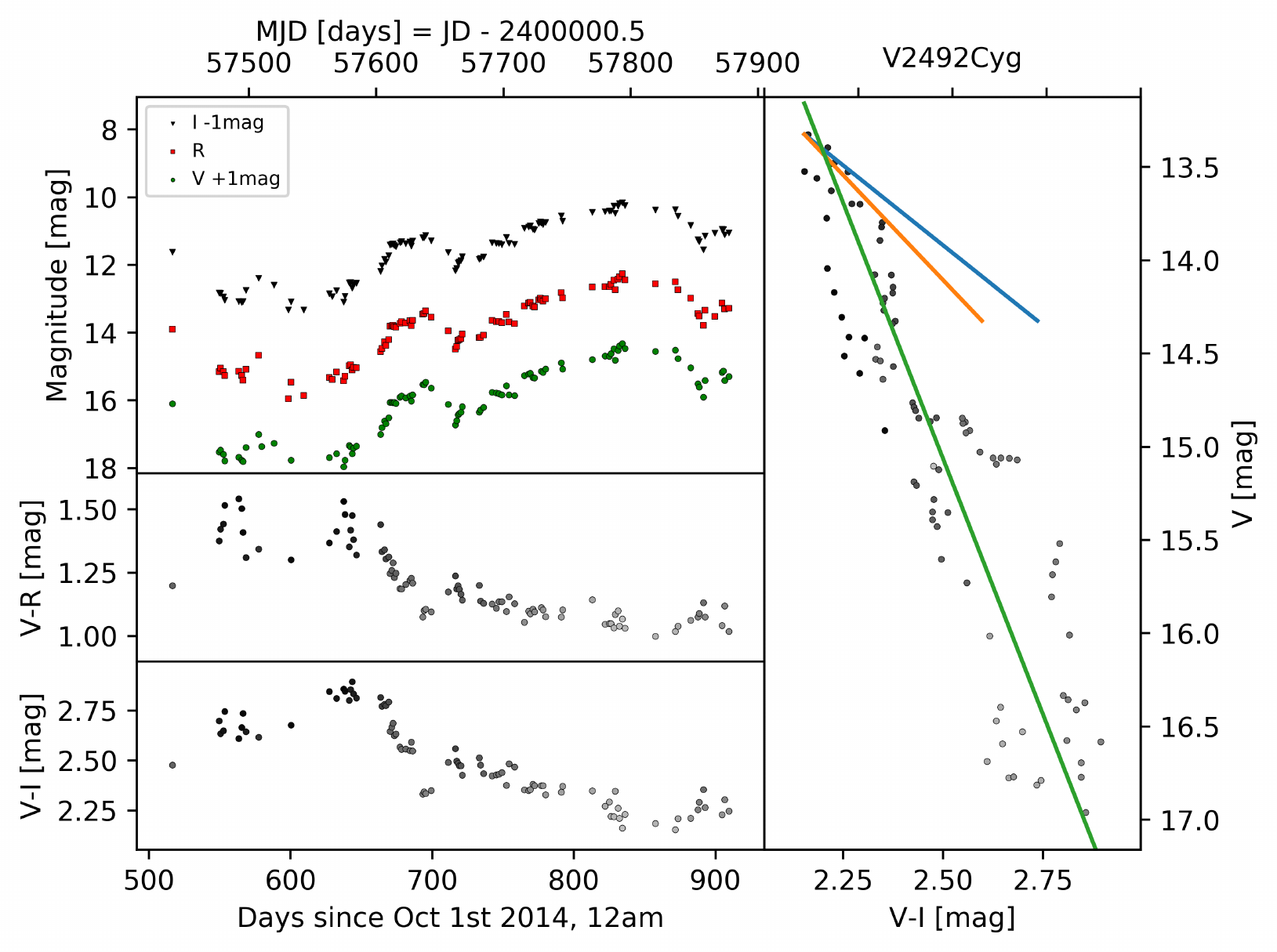} 
\caption{Light-curve of the star V\,2492\,Cyg. In the top left panel we show the instrumental magnitudes (shifted to approximately correspond to the apparent magnitude) for the V, R and I filters. Note that images taken in a different filter have been calibrated into the VRI system. The two panels in the bottom left show the evolution of V-R and V-I colour of the source. The colour of the data-points represents the V-Band magnitude. Black symbols correspond to the faintest magnitudes and light-gray symbols to brighter magnitudes. The right panel shows the position of all observations in the V-I vs V colour-magnitude diagram. Here the colour of the symbols corresponds to the time of the observations. The darkest points correspond to the most recent data. The green line is the line of the best linear fit to the data after removing outliers. The red and blue line indicate extinction vectors for $A_V = 1$\,mag for $R_V = 3.1$ (blue) and $R_V = 5.0$ (red).\label{V2492Cyg}}
\end{figure}

\subsection{Discussion of selected individual Sources}

In this section we discuss three selected objects -- V\,2492\,Cyg, V\,350\,Cep, 2MASS\,J21383981$+$5708470 in more detail. The determined properties for all individual sources are listed in Table\,\ref{yso_data_table} in the Appendix. 

\subsubsection{V\,2492\,Cyg in IC\,5070}

The light-curve of V\,2492\,Cyg is shown in Fig.\,\ref{V2492Cyg}. The source is also known as IRAS\,20496$+$4354, WISE\,J205126.22$+$440523.8 or PTF\,10\,nvg. On 2016-02-22 it was also published as Gaia\,16\,aft. In Table\,\ref{yso_data_table} the source is called R11\,T1\,205126.22$+$440523.7, as this is the designation in the YSO list from \citet{2011ApJS..193...25R}, which we used to select objects in IC\,5070.

The object has been studied by numerous authors in the past, e.g. \citet{2011ApJ...733L...8K, 2011AJ....141..196A, 2011AJ....141...40C, 2011A&A...527A.133K, 2013AJ....145...59H, 2013A&A...551A..62K, 2013MNRAS.430.2910S, 2017arXiv171008151G}. It also was subject of many Astronomer's Telegrams, e.g. \citet{2015ATel.7436....1A, 2017ATel10183....1M, 2017ATel10170....1I, 2017ATel10259....1F}. The object is considered an embedded Class\,I protostar that shows stochastic high amplitude variations in brightness. This variability is driven by changes in accretion rate as well as extinction (up to $\Delta A_V$\,=\,30\,mag), i.e. restructuring of the inner disk. Near infrared spectra are similar to McNeil's Nebula (V\,1647\,Ori) and the source might be considered as a young, embedded analogue to UX\,Ori type stars. Typical variations in brightness occur over timescales of months to years, with a $\sim$220\,day quasi-periodic signal reported in \citet{2013AJ....145...59H}.

All of the data in our light-curve has been included in \citet{2017ATel10259....1F}. There is no indication of the 220\,day quasi-periodic signal identified in \citet{2013AJ....145...59H}. Instead, after a minimum brightness between April and July 2016, the object has steadily increased its brightness to a maximum at the end of January 2017. The object then declined in brightness and the total variations observed in our data are $\Delta$V\,$\approx$\,4\,mag. The colour-magnitude diagram indicates that the variations are consistent with changing extinction from larger dust grains. However, there are large and systematic deviations from this simplistic picture, which indicate changes in accretion rate as well as changes in the properties of the occulting material.

\begin{figure}
\centering
\includegraphics[angle=0,width=\columnwidth]{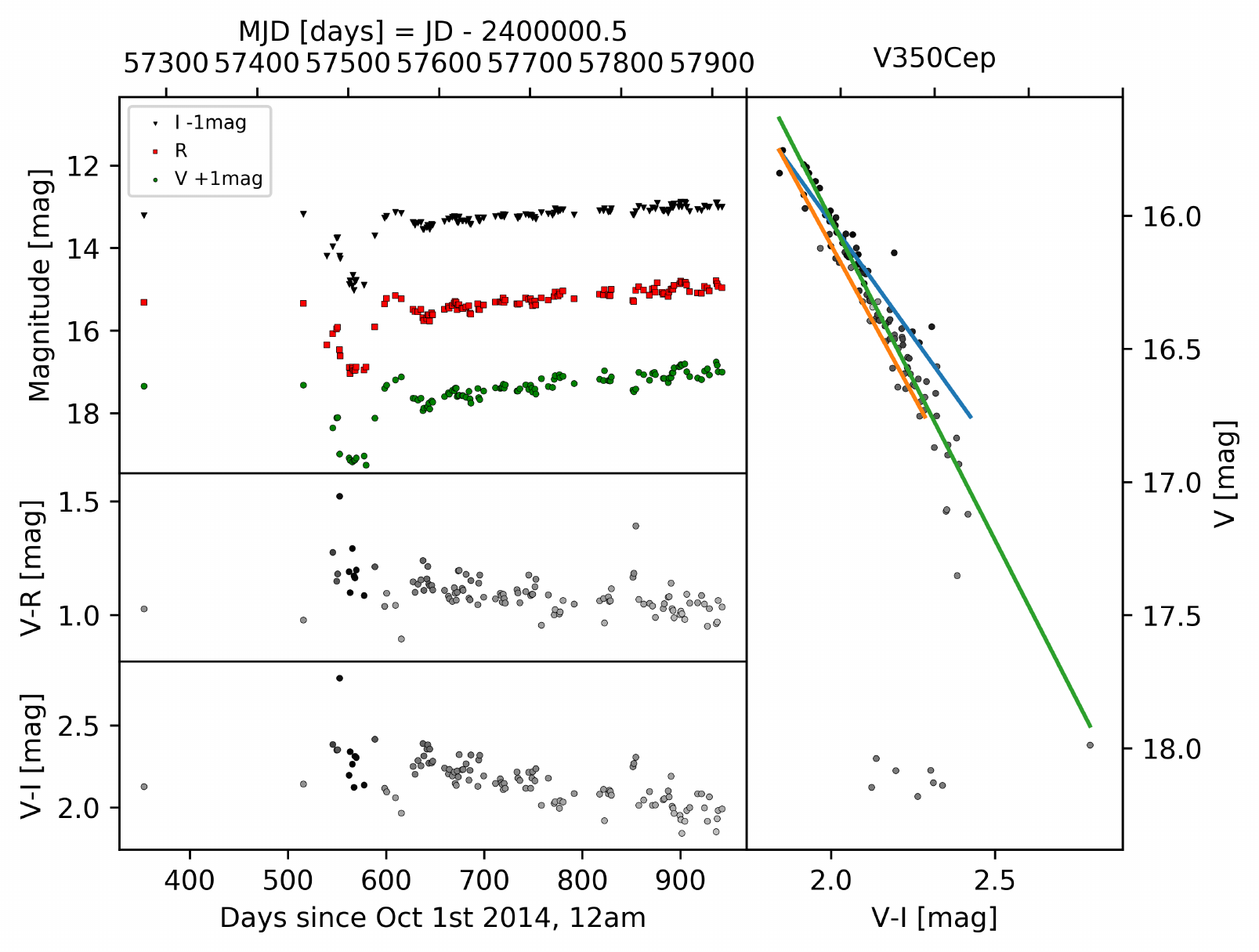} 
\caption{As Fig.\,\ref{V2492Cyg} but for the object V\,350\,Cep \label{V350Cep}}
\end{figure}

\subsubsection{V\,350\,Cep in NGC\,7129}

We show the light-curve of V\,350\,Cep in Fig.\,\ref{V350Cep}. The source is also known as 2MASS\,J21430000$+$6611279, NGC\,7129\,IRS\,1 or NGC\,7129\,FIR\,3. In Table\,\ref{yso_data_table} the source is called HBC$\_$732, as this is the designation in the YSO list from \citet{2009A&A...507..227S}, which we used to select objects in NGC\,7129.

The long term light-curve of this object \citep{2014RAA....14.1264I} shows an about 5\,mag increase in the Blue/pg magnitude between 1965 and 1975, and the source has remained at this level since with typical variations of about 0.5\,mag. The data presented in \citet{2014RAA....14.1264I} suggest a 'dip' of up to 1\,mag in V for several months during 2009. The source has been interpreted as both FU-Ori and EX-Lup (EX-Or) candidate by several authors \citep{2015AJ....149..200D, 1999Ap.....42..121M, 1994A&A...281..864M}. Typical FU-Ori spectral features are missing while EX-Or features are present, however the duration of the burst of now approximately 45\,yrs seems atypical for an EX-Or. 

On 2016-04-23 the object was reported as Gaia\,16\,alt. It showed two Gaia data-points clearly below its usual brightness. This event was also reported by \citet{2017BlgAJ..27...75S} who covered the occultation event with 7 BVRI data-points. Our light-curve in Fig.\,\ref{V350Cep} shows the event at an even higher time resolution. In total 13 VRI measurements are taken during the  occultation event. When combining our data with \citet{2017BlgAJ..27...75S} we find that the object has been detected in occultation between 2016-04-12 and 2016-05-25. Considering the closest data-points outside the occultation we estimate a duration of 51\,--\,83\,days for the event. The light-curve shows that there are at least two 'dips' with an intermittent maximum on 2016-04-17. 

After the occultation the object returned to maximum brightness, just to drop again by about 0.5\,mag in V, from which it has then recovered steadily. We find the depth of the minimum to be $\Delta$V\,=\,2.05\,mag, $\Delta$R\,=\,1.80\,mag and $\Delta$I\,=\,1.90\,mag. These are slightly deeper than reported in \citet{2017BlgAJ..27...75S}. The colour-magnitude plot shows that the first magnitude of decrease during the occultation, as well as the steady recovery of the flux after it, are in good agreement with a variation caused by changes in extinction due to normal ISM dust grains. During the deeper part of the occultation, the change in brightness turns completely free of colour change. Hence, the denser part of the occulting material responsible for this seems to have consisted of larger dust grains or the obscuring material was optically thick.

\begin{figure}
\centering
\includegraphics[angle=0,width=\columnwidth]{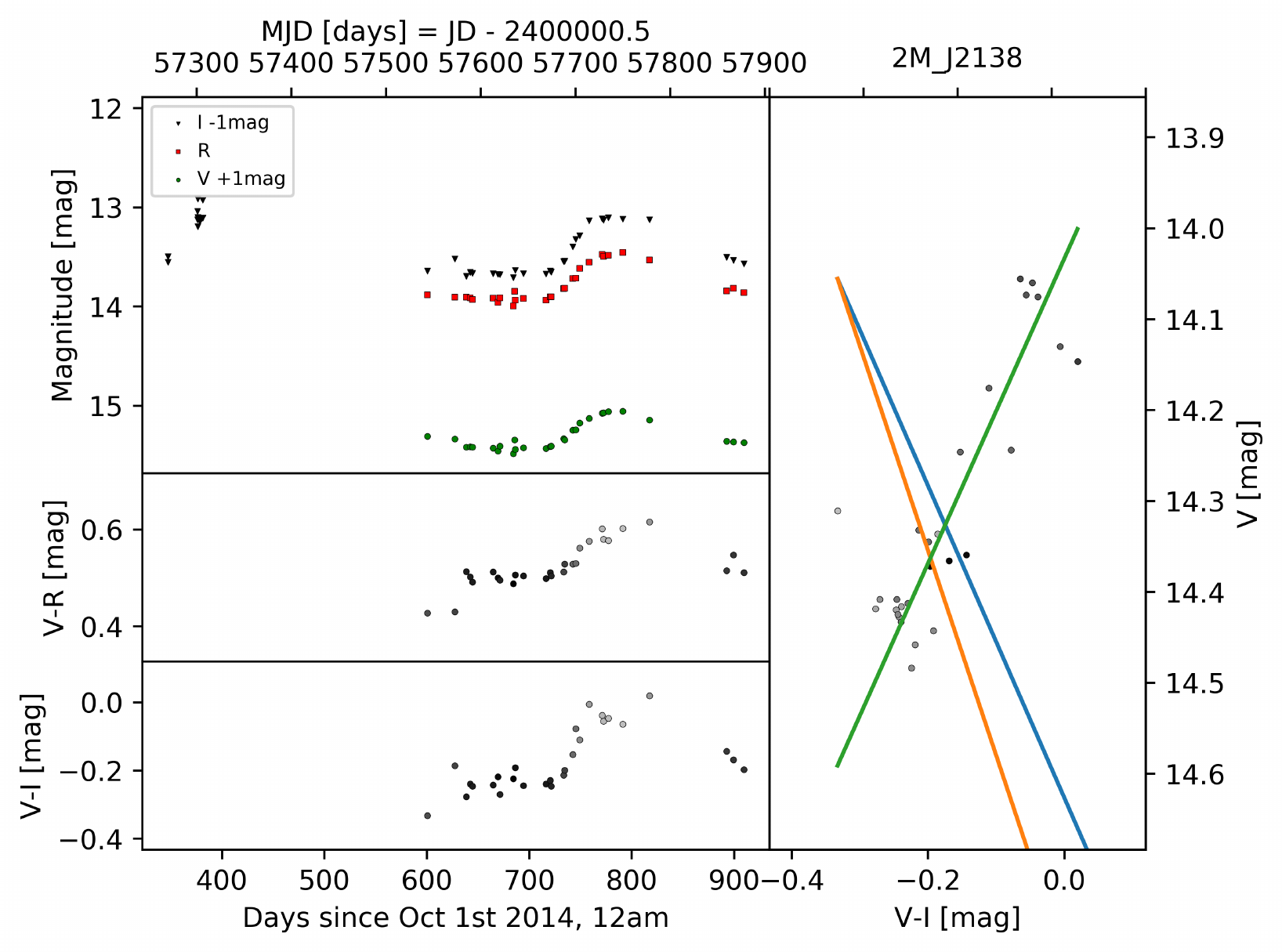} 
\caption{As Fig.\,\ref{V2492Cyg} but for the object 2MASS\,J21383981$+$5708470 \label{J2138}}
\end{figure}

\subsubsection{2MASS\,J21383981$+$5708470 in IC\,1396\,A}

The light-curve of the source 2MASS\,J21383981$+$5708470, also known as $[$NSW2012$]$\,284 is shown in Fig.\,\ref{J2138}. This star is listed as Emission Line star in SIMBAD and has been identified in \citet{2012AJ....143...61N}. In Table\,\ref{yso_data_table} the source is called Variable$\_$V$\_$324.665863$\_$57.146301, as it was added as a variable source due to its large Stetson index in the V-Band and was not included in the YSO candidate list from \citet{2005AJ....130..188S}, which we used to select objects in IC\,1396\,A.

The object shows a smooth brightness increase over about 150 days in all three filters. At the peak the brightness increase in V is about 0.4\,mag, while in the I-Band the star got about 0.6\,mag brighter. The Stetson index of the source is 1.36 and the asymmetry index -0.74, i.e. the object is classified as burster. What is highly unusual about this source is the change in colour towards red during the outburst. The angle in the V vs. V-I diagram is -59\dg. This is the only source in the entire sample with such a behaviour and consequently the object has been made part of the outlier group G\,7 by the hierarchical clustering algorithm. The older I-Band (only) data suggests that there might have been another such burst about 600\,days prior to the one covered by our survey. But only a small part of this potential burst is covered. If this is a periodic behaviour, then the next burst should occur at about MJD\,=\,58300\,d -- which is early 2018. The nature of this source is unclear. Usually objects in outburst are bluer in the bright state. Hence, the object has either an unusual burst, or we see only scattered light from this source.

\section{Conclusions}

We are describing and presenting the first results of our optical survey of nearby clusters and star forming regions with small telescopes. All observations are obtained with the University of Kent's 17\arcsec\ Beacon Observatory or as part of the HOYS-CAPS citizen science project. In this paper we present the analysis of variable stars in eight target fields for which V, R and I-Band data has been taken over a period of about two years. 

In our dataset we have identified 466 variable stars, 413 of which are confirmed YSOs, based on the Stetson index of their light-curves. For all objects light-curve properties such as the the asymmetry metric and slope ($\alpha$) in the V-I vs. V colour-magnitude diagram are determined. This sample is one of the largest samples of variable YSOs in the northern hemisphere with multi colour observations of such a time span and cadence, with additional archival multi-wavelength data and which is well suited for follow up observations. We find that the number of protostars in our sample is negligible, while about 65\,\% of the objects are CTTSs and about 35\,\% are sources with transition or depleted disks. 

A detailed investigation of the asymmetry index $M$ and the $\alpha$-values of the light-curves has been performed. We can use the $\alpha$-value to characterise the most likely mechanism responsible for the variability in the stars. These range from occultations by material composed of large dust grains or optically thick material (including eclipses) to occultations by material consistent with normal ISM dust to changes in mass accretion rates or dark/bright spots. Dipper light-curves consistent with changes in accretion rates or spots are rare, there are virtually no burster objects that are consistent with occultations or eclipses. Burster light-curves also show by far the highest variability (of the order of 1.1\,mag in V), followed by dippers and eclipsing objects (0.7\,mag), and  the symmetric light-curves have the lowest variations (0.5\,mag).

We used a hierarchical clustering algorithm to identify groups of YSOs with similar light-curve properties. This clustering algorithm allows us to identify the most unusual objects in our sample. We further find that clustering results in groups of YSOs that correspond to dippers, bursters or symmetric light-curves and they also have different $\alpha$-values. Thus, grouping variable YSOs via hierarchical clustering using their light-curve properties is a promising approach for future studies of larger samples of YSOs that will become available e.g. from GAIA or the LSST. 

\section*{acknowledgements}

J.\,Campbell-White acknowledges the studentship provided by the University of Kent. S.V.\,Makin acknowledges an SFTC scholarship (1482158). KW acknowledges funding by STFC. K.\,Wiersema thanks Ray Mc\,Erlean and Dipali Thanki for technical support of the UL50 operations. This research made use of Montage. It is funded by the National Science Foundation under Grant Number ACI-1440620, and was previously funded by the National Aeronautics and Space Administration's Earth Science Technology Office, Computation Technologies Project, under Cooperative Agreement Number NCC5-626 between NASA and the California Institute of Technology. We acknowledge ESA Gaia, DPAC and the Photometric Science Alerts Team (http://gsaweb.ast.cam.ac.uk/alerts). 

\bibliographystyle{mn2e}
\bibliography{biblio}
\label{lastpage}

\clearpage
\newpage

\begin{appendix}

\section{Data of Individual Sources}

\onecolumn

\begin{landscape}
\renewcommand{\tabcolsep}{3pt}
\setlength\LTcapwidth{\textheight}

\begin{longtable}{|c|c|c|c|c|c|c|c|c|c|c|c|c|c|c|c|}

\caption{\label{yso_data_table} This table lists the individual properties of all stars investigated in the
papers. We list the catalogue name of the source, the name of the cluster/region it is in, the Right Ascention and
Declination (J2000), the number of points in the lightcurve used to determine the slope $\alpha$ in the V vs V-I
diagram, the slope $\alpha$ of a linear fit in the V vs V-I; the $rms$ of the datapoints from the fit, the Asymmetry
index $M$, the Stetson index $S$, the group in the dendrogram the source is associated with, the range of magnitudes and
the $rms$ of the V, R and I-Band lightcurve.} \\

\hline
Source Name & Region & RA & DEC & N$_{\rm var}$ & $\alpha$ & $rms$ & $M$ & $S$ & G & $\Delta{\rm V}$ & $\sigma_{\rm V}$ & $\Delta{\rm R}$ & $\sigma_{\rm R}$ & $\Delta{\rm I}$ & $\sigma_{\rm I}$  \\ 
 &  & \multicolumn{2}{c|}{(J2000) [deg]} &  & [deg] & [mag] &  &  &  & [mag] & [mag] & [mag] & [mag] & [mag] & [mag]  \\ \hline
\endfirsthead

\multicolumn{16}{c}%
{{\bfseries \tablename\ \thetable{} -- continued from previous page}} \\
\hline
Source Name & Region & RA & DEC & N$_{\rm var}$ & $\alpha$ & $rms$ & $M$ & $S$ & G & $\Delta{\rm V}$ & $\sigma_{\rm V}$ & $\Delta{\rm R}$ & $\sigma_{\rm R}$ & $\Delta{\rm I}$ & $\sigma_{\rm I}$  \\ 
 &  & \multicolumn{2}{c|}{(J2000) [deg]} &  & [deg] & [mag] &  &  &  & [mag] & [mag] & [mag] & [mag] & [mag] & [mag]  \\ \hline
\endhead

\hline \multicolumn{16}{|r|}{{Continued on next page}} \\ \hline
\endfoot

\hline
\endlastfoot

2MASS$\_$J03281101$+$3117292 & NGC1333 &    52.045917 &     31.29147 &  19 & 44.0 & 0.013 & -0.24 &  1.72 & 6 &  0.80 &  0.19 &  0.14 &  0.04 &  0.05 &  0.02 \\ 
2MASS$\_$J03283692$+$3117353 & NGC1333 &    52.153833 &     31.29317 &  18 & 45.4 & 0.047 &  0.04 &  2.09 & 6 &  0.96 &  0.24 &  0.31 &  0.08 &  0.24 &  0.06 \\ 
2MASS$\_$J03283692$+$3117353 & NGC1333 &    52.153833 &     31.29317 &  19 & 46.0 & 0.046 &  0.28 &  1.66 & 5 &  0.60 &  0.18 &  0.32 &  0.07 &  0.24 &  0.06 \\ 
2MASS$\_$J03285101$+$3118184 & NGC1333 &    52.212542 &     31.30514 &  23 & 72.1 & 0.101 &  0.90 &  2.33 & 1 &  1.21 &  0.26 &  1.06 &  0.23 &  1.28 &  0.26 \\ 
2MASS$\_$J03285216$+$3122453 & NGC1333 &    52.217333 &     31.37925 &  24 & 71.3 & 0.048 &  0.91 &  2.37 & 1 &  0.79 &  0.22 &  0.73 &  0.20 &  0.57 &  0.16 \\ 
2MASS$\_$J03285741$+$3119505 & NGC1333 &    52.239208 &     31.33072 &  17 & 46.6 & 0.021 &  0.87 &  1.02 & 3 &  0.71 &  0.15 &  0.23 &  0.06 &  0.13 &  0.03 \\ 
2MASS$\_$J03285954$+$3121467 & NGC1333 &    52.248125 &     31.36300 &  23 & 78.5 & 0.107 & -0.06 &  6.62 & 6 &  3.09 &  0.75 &  2.74 &  0.66 &  2.49 &  0.62 \\ 
2MASS$\_$J03290386$+$3121487 & NGC1333 &    52.266083 &     31.36353 &  25 & 50.7 & 0.085 &  0.39 &  1.23 & 3 &  0.55 &  0.14 &  0.53 &  0.15 &  0.44 &  0.12 \\
\end{longtable}

\end{landscape}

\twocolumn

\end{appendix}

\end{document}